\documentclass[a4paper, 12pt]{article}
\usepackage[utf8]{inputenc}
\usepackage{color}
\usepackage{hyperref} 
 \usepackage{latexsym}
  \usepackage{amssymb}
  \usepackage{graphicx}
   \DeclareGraphicsExtensions{.jpg .png .gif .pdf .pdf}
\usepackage[T1]{fontenc}
 \usepackage[english]{babel}
  \usepackage{}
  \usepackage{amstext}
  \usepackage{amsmath}
  \usepackage{times} 
  \usepackage{amsfonts}
 \date{}

\title{Nonlinear dynamics of plasma oscillations modeled by a 
forced modified Van der Pol-Duffing oscillator }
\author{C. H. Miwadinou\footnote{clement.miwadinou@imsp-uac.org, hodevewan@yahoo.fr}, 
L. A. Hinvi\footnote{laurent.hinvi@imsp-uac.org} ,  
A. V. Monwanou\footnote{movins2008@yahoo.fr} and \\
J. B. Chabi Orou\footnote{Author to whom correspondence should be addressed: jchabi@yahoo.fr}}

\begin{document}

\maketitle

\begin{abstract}
This paper considers nonlinear dynamics of plasma oscillations modeled by 
a forced modified Van der Pol-Duffing oscillator.
These plasma oscillations are described by a nonlinear differential equation of the form
$ \ddot{x}+ \epsilon \left( 1 +{x}^{2} \right){\dot{x}} + x+\epsilon \alpha {x}{\dot{x}} +
{\beta}x^{2}+\gamma x^{3}= F\cos{\Omega t}.$ The amplitudes of the forced harmonic, 
superharmonic and subharmonic oscillatory states are obtained using the harmonic balance 
technique and the multiple time scales methods. Admissible values of the amplitude of 
the external strength are derived. Bifurcation sequences displayed by the model for each
type of oscillatory states are performed numerically through the fourth order Runge- Kutta scheme.
\end{abstract}

\section{Introduction}
Many problems in physics,  chemistry,  biology,  etc.,  are related to nonlinear
self-excited oscillators \cite{24}. Intrests according to plasma oscillations are due to their potential applications. Indeed,  
radio-wave propagation in the ionosphere was an early stimulus for the development of the theory of plasma. Nowadays,  plasma processing
is viewed as critical technology in a large number of industries. It is also important in other sectors such as biomedecine,  automobiles, 
defence,  aerospace,  optics,  solar energy,  telecommunications,  textiles,  papers,  polymers  and waste managment \cite{25}
 For example,  in particular, the nonlinear description of plasma oscillations
is of interest as a result of its importance to the semiconductor industry\cite{8}, \cite{9}. 
Experiments suggest that some plasma behavior is approximately described 
by anharmonic oscillations. Thus,  it has been shown experimentally \cite{10} and theorytically \cite{11}  that in
plasma physics,  the electron beam surfaces,  the Tonks-Datter resonances of mercury vapor and
low frequency ion sound waves,  oscillations are described by the following anharmonic equations 
 (\ref{eq0}) or  (\ref{eq01}) \cite{12}.

\begin{eqnarray}
 \ddot{x}+\omega_0^2x+{\beta}x^{2}+
  \gamma x^{3}&=& F\cos{\Omega t};\label{eq0}
\end{eqnarray}

\begin{eqnarray}
 \ddot{x}+ \epsilon \left( 1 +{x}^{2} \right){\dot{x}}
   +\omega_0^2 x +{\beta}x^{2}+
  \gamma x^{3}&=& F\cos{\Omega t}.\label{eq01}
\end{eqnarray}

In these equations,  $\omega_0 $ and $\Omega $ are respectively the natural and external
frequencies. $F$ stands for the amplitude of 
the external excitation while $\beta,  \gamma$ and $\epsilon$ are the quadratic, 
cubic nonlinearities and dampind parameters.  
In \cite{11},  the authors studied the regular and chaotic behaviors of plasma oscillations.
In \cite{12},  the authors studied  with a rigorous 
theoretical consideration that through a method based on the harmonic-balance formalism 
the amplitude of the forced harmonic oscillations states. It is 
found also the resonance states and admissible values of $F$ and bifurcation structures
are numerically found. They confirmed this prediction 
with numerical simulations  and the important effects of differents parameters.  

In the present paper,  we concentrate our studies on the model and equation of motion,  the
resonant states,  the chaotic behavior. Through these studies,  we found the effects of differents
parameters in general and in particular on the effect of the hybride quadratic parameter $\alpha$ which shown
the difference between this equation and anharmonic equation which obtained in \cite{12}.
  
The paper is structured as follows: Section $2$ gives the model and equation of nonlinear dynamics of plasma oscillations.
Section $3$ gives an analytical treatment of equation (\ref{eq20}). Amplitude of the forced harmonic oscillatory 
states 
is obtained with harmonic-balance method \cite{1}. Section 4 investigate using multiple time-scales
method \cite{2} the resonant cases and the stability conditions are found by the perturbation method 
\cite{1}. The section ends evaluates  bifurcation and chaotic behavior by numerically
simulations of equation (\ref{eq20}). Section 6  deals with conclusions.

\section{MODEL AND EQUATION OF MOTION}

We consider the two-fluid model which treats the plasma as two inter penetrating conducting fluids.
This model consists of a set of fluid equations for the electrons and ions plus the
complete set of Maxwell’s equations. Such a model has been source of growing interest for researchers
for many years \cite{13} to \cite{15} and nowadays,  it remains an interesting task 
because of its potential applications\cite{16} to \cite{20}.
The two-fluid plasma system is applicable in many areas where high density
plasmas interact with high frequency electromagnetic waves. A few examples include application
to helicon thrusters,  electron cyclotron resonance plasma sources and plasmas resulting from
strong explosions in the atmosphere. In the electrostatic situation it also has possible 
application in modeling discharge cathode plasma sources which are important in ion
thrusters and hall thrusters. The Eulerian equations of motion in electric and magnetic fields $E$, 
$B$ are given as follows \cite{21}:

\begin{eqnarray}
 n_{0}M_{\alpha}\frac{dv_{\alpha}}{d\tau}&=& n_{0}e\left( E+ v_{\alpha}\wedge B_{0}-\eta J\right) 
 -{\triangledown}{P_{\alpha}},  \label{eq1}\\
 \frac{\partial{n_{\alpha}}}{\partial{\tau}} +{\triangledown}.\left(n_{0}v_{\alpha}\right)&=& S
 \label{eq2}\\
   \frac{d}{d\tau}\left(P_{\alpha}{n_{\alpha}}^{-\gamma}\right)&=&0. \label{eq3}
\end{eqnarray}

$S$ is the source term due to ionization or to large amplitude oscillations present in the plasma. 
The suffix ''$\alpha$'' stands for the species label and it will be denoted by $i$ and $e$
respectively for positive ions with charge $+e$ and electron with charge $-e$. $n_{\alpha}$
stands for the density of the species,  $v_{\alpha}$
their velocity,  $P_{\alpha}$ their pressure,  $\gamma $ the usual specific heat ratio and $\eta$ 
the resistive collision which is defined as
\begin{equation}
 \eta=\frac{M\nu_{\alpha}}{n e^{2}}, \label{eq4}
\end{equation}

where $\nu_{\alpha}$ is the collision frequency of the species $\alpha$.
The electric charge density $\rho$ and current $J$ are given by

\begin{equation}
   \rho={\sum}_{\alpha}n_{\alpha}q_{{\alpha}}, 
 J={\sum}_{\alpha}n_{\alpha}q_{{\alpha}}v_{\alpha}.\label{eq5}
\end{equation}
These quantities are the source terms for Maxwell’s equations. In order to deal with small amplitude waves,  
we consider a ”background” situation representing uniform infinite plasma. The
values of $n_\alpha$ ,  $v_{\alpha}$ ,  $P_\alpha$ ,  $E$,  $B$ for this will be denoted by 
$n_{0 {\alpha}}$ ,  etc; however,  here we shall take
$v_\alpha = E = 0$ in the unperturbated state. We then have $J = 0$ and all of 
Eqs.(\ref{eq1})-(\ref{eq5}) are satisfied except Eqs. (\ref{eq5}) which require $\rho = 0$,  hence
\begin{equation}
 {\sum}_{\alpha}n_{\alpha}q_{{\alpha}}=0. \label{eq6}
\end{equation}
For our simple two-species plasma,  that condition of charge neutrality becomes
$$n_{0e} = n_{0i} = n_{0}.$$
We now consider the $M_{e} = 0$ ion-sound instability and introduce perturbations terms which are
denoted by the suffix $1$,  namely
\begin{equation}
 n_{i} = n_{0} + n_{1},   P_{i} = P_{ 0} + P_{1},  B_{i} = B_{ 0} + B_{1}\label{eq7}
\end{equation}
Let us note that for other variables which vanish at the unperturbed state,  labels $0$ and $1$
are not necessary. $n_{1}$ is considered as a perturbed time varying function of $z$.
We then insert the expressions (\ref{eq7}) into Eqs. (\ref{eq1})-(\ref{eq3}) 
and after all of the second order perturbation terms have been discarded, 
we obtain the following equations.

\begin{eqnarray}
 n_{0}M_{i}\frac{dv_{i}}{d\tau}&=& n_{0}e\left( E+ v_{i}\wedge B_{0}-{\eta} J\right) 
 -{\triangledown}{P_{1}},  \label{eq8}  
\end{eqnarray}
\begin{eqnarray}
 \frac{\partial{n_{1}}}{\partial{\tau}} +{\triangledown}.\left(n_{0}v_{i}\right)&=& S\label{eq9}\\
 \frac{P_{1}}{p_{0}}&=&\gamma \frac{n_{1}}{n_{0}} \label{eq10}
\end{eqnarray}
Since in this case of the ion-sound instability under consideration,  only spatial variation of 
the form $e^{-jk_{z} z}$ ($j$ is a complex number) needs to be considered \cite{14}.
Here,  $k_{z}$ stands for the wave number in the $z$ direction.
In dealing with Eq.(\ref{eq3}) and taking each species to be a perfect gas with unperturbed temperature
$T$ (which could be different for each species),  we have 
\begin{eqnarray}
 P_{0}&=&n_{0}k_{B}T_{i}, \label{eq11}
\end{eqnarray}
where $k_{B}$ is Boltzmann’s constant and Eq. (\ref{eq8}) can be rewritten as follows:

\begin{eqnarray}
 n_{0}M_{i}\frac{dv_{i}}{d\tau}&=& n_{0}e\left( E+ v_{i}\wedge B_{0}-{\eta} J\right) 
 -{\gamma }k_{B}T_{i}{\triangledown}n_{1}. \label{eq12}  
\end{eqnarray}
To investigate the two-fluid model,  we assume that
\begin{eqnarray}
 E&=& -{\triangledown}{\phi}, \label{eq13}
\end{eqnarray}
 where ${\phi}$ is the corresponding potential
and consider the Boltzmann distribution equation of electron given as follow:
\begin{eqnarray}
 \frac{n_{1}}{n_{0}}&=&\frac{e\phi}{k_{B}T_{e}}\label{eq14}
\end{eqnarray}
By considering
\begin{eqnarray}
\frac{dv_{i}}{d\tau}&=&\frac{\partial{v_{i}}}{\partial{\tau}}+\left(v_{i}.
{\triangledown}\right)v_{i}, \label{eq15}
\end{eqnarray}
eliminating $v_{i}$ between Eqs.(\ref{eq2}) and (\ref{eq12}),  we obtain after
some algebraic manipulations the following equation

\begin{eqnarray}
 \frac{d^{2}{n_{1}}}{{d{\tau}}^{2}}  -\frac {k_{B}}{M_{i}}\left(T_{e}+
 {\gamma }T_{i}\right){\triangledown}^{2} {n_{1}} +  
 \frac{ n_{0}e}{M_{i}}{\triangledown}.\left(v_{i}\wedge B_{0}\right)+\nonumber\\
 -{\nu}_{i}\left(S-\frac{\partial{n_{1}}}{\partial{\tau}}\right)-\frac{dS}{d{\tau}}&=&0. \label{eq16}
\end{eqnarray}
Keen and Fletcher\cite{13}, \cite{14} and Hsuan \cite{21} in the early seventies showed that 
from thermodynamics argument,  the source term is a function of density $n_{1}$.

Here,  the source is taken to be of the following form
\begin{eqnarray}
 S&=&-\lambda {n_{1}}^{2}-\mu{n_{1}}^{3} \label{eq17}, 
\end{eqnarray}
By assuming that the model is influenced by an external sinusoidal excitation 
$F_{0} \cos {\omega \tau}$ ,  Eq.(\ref{eq16})becomes

\begin{eqnarray}
 \frac{d^{2}{n_{1}}}{{d{\tau}}^{2}}+ \left( {\nu}_{i} +
 2\lambda {n_{1}}+3\mu{n_{1}}^{2} \right) \frac{d{n_{1}}}{d{\tau}}
 +\frac {k_{B}}{M_{i}}\left(T_{e}+{\gamma }T_{i}\right){\vec{k}}^{2} {n_{1}} \nonumber\\
-j\frac{ n_{0}e}{M_{i}}{\vec{k}}.\left(v_{i}\wedge B_{0}\right)  
 +{\nu}_{i}\left(\lambda {n_{1}}^{2}+\mu{n_{1}}^{3}\right) &=&F_{0} \cos {\omega \tau}, \label{eq18}
\end{eqnarray}
where ${\triangledown}=-j\vec{k}.$ 

If one considers the slab geometry configuration for which density varies in the $x$-direction and
the $z$-axis coincides with the magnetic field direction,  (\ref{eq18}) takes
the following expression

\begin{eqnarray}
   \frac{d^{2}{n_{1}}}{{d{\tau}}^{2}}+ \left( {\nu}_{i} +
   2\lambda {n_{1}}+3\mu{n_{1}}^{2} \right) \frac{d{n_{1}}}{d{\tau}}
   +{\omega_{0}}^{2}n_{1}+{\nu_{i}}\left( {\lambda} n_{1}^{2}+
   {\mu} {n_{1}}^{3}\right)&=&\nonumber\\F_{0}\cos{\omega \tau}; \label{eq19}
\end{eqnarray}
Following the rescaling\\
$ {\omega_{0}}=k_{z}C_{K}$,  $C_{K}=\sqrt{k_{B}\left(\frac{T_{e}+\gamma T_{i}}{M_{i}}\right)}$, 
$t=\tau {\omega_{0}}$,  
$n_{1}=\left( \frac{\nu_{i}}{3\mu}\right)^{1/2}x=Ax$,  
$F=\frac{F_{0}}{A{\omega_{0}}^{2}}$,  $\Omega=\frac{\omega}{\omega_{0}}$.

It comes that the system is governed by the following nonlinear second order differential 
anharmonic equation
\begin{eqnarray}
 \ddot{x}+ \epsilon \left( 1 +{x}^{2} \right){\dot{x}}
   +{x}+ \epsilon\alpha {x}{\dot{x}} +{\beta}x^{2}+
  \gamma x^{3}&=& F\cos{\Omega t}; \label{eq20}
\end{eqnarray}
with:
$\epsilon=\frac{\nu_{i}}{\omega_{0}}$,  
$\alpha=\frac{2\lambda}{{\nu}_{i}}\left( \frac{\nu_{i}}{3\mu}\right)^{{1}/{2}}$, 
$\beta=\frac{{\lambda}{\nu}_{i} }{{\omega_{0}}^{2}}A
 = \frac{{\lambda}{\nu}_{i} }{{\omega_{0}}^{2}}\left(\frac{\nu_{i}}{3\mu}\right)^{{1}/{2}}$, 
$ \gamma=\frac{{\mu}{\nu}_{i} }{{\omega_{0}}^{2}}A^{2}
=\frac{{\mu}{\nu}_{i} }{{\omega_{0}}^{2}} \frac{\nu_{i}}{3\mu}$.

This equation (\ref{eq20}) is a forced modified Van der Pol-Duffing oscillator 
equation. There is the equation that modeled the non linears dynamics oscillators  
of the plasma. There are several physical mechanisms that could mimic the driven force $F$. For example,  the
transport of dust particles into plasma is proportional to the dust charge and as well as to the
coagulation of small particles into larger ones since charged particles attract or repel each other
through the coulomb potential. But,  considering the fact that ultraviolet light can extract electrons
from materials by photo-detachment,  such a light can be used as an external force to control the
charge on a dust particle \cite{22}. Such forcing terms could also be mimicked through an externally
applied electric field that supplies the system with an external drive\cite{16}. 
In particular when the friction term vanish ($\epsilon=0$),  
then the equation  reduces to forced modified Duffing oscillator equation 
studies by H. G. Engieu Kadji and al.($2005$). 
When $\epsilon \neq0$ and $\alpha=0$,  the equation (\ref{eq20}) reduces to
an anharmonic oscillator, 
study two yields ago by H. G. Engieu Kadji and al.($2007$)\cite{12}.
It should be quoted that the coefficient of the dissipation term plays a key role
on how a limit cycle is born and how it dies. Additionally,  non
dissipative plasma that corresponds to the ideal case has been mainly considered in the two-fluid
stationary states studies for many decades. But,  since the plasma is dissipative and externally
driven in realistic experimental situations,  it is of interest to elaborate a formalism for investigating
the driven dissipative two-fluid model in order to forecast theoretical results closer to those of the
experiments. We  will also  see the condition of Hopf bifurcation apparition.
We  will finaly  study the  transition to chaos of the system.
\section{AMPLITUDE OF THE FORCED HARMONIC OSCILLATORY STATES}
Assuming that the fundamental component of the solution and the external 
excitation have the same period,  the amplitude of harmonic oscillations can be
tackled using the harmonic balance method \cite{1}. For this purpose,  we express
its solutions as
\begin{eqnarray}
 x &=& A \cos\left(\Omega-\psi\right)t  +\xi \label{eq21}
\end{eqnarray}
where $A$ represents the amplitude of the oscillations and $\xi$
a constant. Inserting this solution (\ref{eq21})in (\ref{eq20})
 and equating the constants and the coefficients of
 $\sin{\omega}t$ and $\cos{ \omega t}$,  we have
 \begin{eqnarray}
  \left[(1-{\Omega}^{2})A+ 2{\beta}A{\xi }
  +\gamma(\frac{3}{4}A^{3}+3A{\xi}^{2} )\right]^{2}
  &=&\nonumber\\-{\epsilon}^{2}\left[A\Omega(1+\alpha{\xi }+{\xi }^{2})+
  \frac{1}{4}A^{3}\Omega\right]^{2} +F^{2}, \label{eq22}
 \end{eqnarray}
 \begin{eqnarray}
  \left(1 +\frac{3}{2}\gamma A^{2}\right){\xi} 
  +\frac{1}{2}\beta A^{2}+\beta {\xi}^{2} +\gamma {\xi}^{3}&=&0. \label{eq23}
 \end{eqnarray}
 If it is assumed that $|{\xi}|\ll |A|$,  i.e that shift in $x = 0$ 
 is small compared to the amplitude \cite{8} ,  then ${\xi}^{2}$ 
and ${\xi}^{3}$  terms in (\ref{eq23}) can be neglected and one obtains
 
 \begin{eqnarray}
{\xi} &=&- \frac{\beta A^{2}}{\left(2 +3\gamma A^{2}\right)}. \label{eq24}
 \end{eqnarray}
 Substituting  (\ref{eq24}) into  (\ref{eq22}) leads us to the 
 following nonlinear algebraic equation 
 
  \begin{eqnarray}
 \left[\frac{9}{4} {\gamma}^{2}A^{5}+\left(\frac{9}{2}\gamma 
 -3\gamma {\Omega}^{2}-2{\beta}^{2}\right)A^{3} +2(1-{\Omega}^{2})^{2}A\right]^{2}\nonumber\\
 +{\epsilon}^{2}{\Omega}^{2}\left[A\left(1+\frac{1}{4}A^{2}\right) (2+3\gamma A^{2})
 -{\alpha}{\beta}A^{3}\right]^{2}\nonumber\\
 -F^{2}\left(2+3\gamma A^{2}\right)^{2}&=&0\label{eq25}
 \end{eqnarray}
 $\Longleftrightarrow$
 \begin{eqnarray}
  \frac{9}{16}\left(9{\gamma}^{2}+{\epsilon}^{2}{\Omega}^{2}\right){\gamma}^{2}A^{10}+\nonumber\\
  +\left[\frac{9}{2}{\gamma}^{2}\left(\frac{9}{2}\gamma-3\gamma {\Omega}^{2}-2{\beta}^{2}\right)
  +\frac{9}{2}{\epsilon}^{2}{\Omega}^{2}{\gamma}^{2}+\frac{3}{8}{\epsilon}^{2}\gamma
  -\frac{3}{2}\alpha \beta \gamma {\epsilon}^{2}{\Omega}\right]{A}^{8}+\nonumber\\
  +\left[9{\gamma}^{2}\left(1-{\Omega}^{2}\right) +\left(\frac{9}{2}\gamma -3\gamma {\Omega}^{2}
  -2{\beta}^{2}\right)^{2}+9{\epsilon}^{2}{\gamma}^{2}{\Omega}^{2}+
  3{\epsilon}^{2}{\gamma}{\Omega}^{2}\right]A^{6}+\nonumber\\
  +\left[\frac{1}{4}{\epsilon}^{2}{\Omega}^{2} -6\alpha \beta \gamma{\epsilon}^{2}{\Omega}^{2}
  -\alpha \beta {\epsilon}^{2}{\Omega}^{2}+
  {\epsilon}^{2} {\alpha}^{2} {\beta}^{2}{\Omega}^{2}\right]A^{6}+\nonumber\\
 \left[4\left(1-{\Omega}^{2}\right)\left(\frac{9}{2}\gamma
 -3\gamma{\Omega}^{2}-2{\beta}^{2}\right)\right]A^{4}+\nonumber\\
 +\left[6{\epsilon}^{2}{\Omega}^{2}+6{\epsilon}^{2}{\Omega}^{2}\gamma
 -4\alpha \beta{\epsilon}^{2}{\Omega}^{2}-9{\gamma}^{2}F^{2}\right]A^{4}+\nonumber\\
 +\left[4\left(1-{\Omega}^{2}\right)^{2}-6\gamma F^{2}\right]A^{2}-4F^{2}&=&0\nonumber\\ \label{eq26}
 \end{eqnarray}

This equation can rewritte as:
\begin{eqnarray}
 A^{10}+P_{1}A^{8}+P_{2}A^{6}+P_{3}A^{4}+P_{4}A^{2}-4F&=&0,  \label{eq27}
\end{eqnarray}
with
\begin{eqnarray}
 P_{1}&=&\frac{\frac{9}{2}{\gamma}^{2}\left(\frac{9}{2}\gamma-3\gamma {\Omega}^{2}-2{\beta}^{2}\right)
  +\frac{9}{2}{\epsilon}^{2}{\Omega}^{2}{\gamma}^{2}+\frac{3}{8}{\epsilon}^{2}\gamma
  -\frac{3}{2}\alpha \beta \gamma {\epsilon}^{2}{\Omega}}
  {\frac{9}{16}\left(9{\gamma}^{2}+{\epsilon}^{2}{\Omega}^{2}\right){\gamma}^{2}}, \label{eq28}\\
  P_{2}&=&\frac{9{\gamma}^{2}\left(1-{\Omega}^{2}\right) +\left(\frac{9}{2}\gamma -3\gamma{\Omega}^{2}
  -2{\beta}^{2}\right)^{2}+9{\epsilon}^{2}{\gamma}^{2}{\Omega}^{2}+
  3{\epsilon}^{2}{\gamma}{\Omega}^{2}}
  { \frac{9}{16}\left(9{\gamma}^{2}+{\epsilon}^{2}{\Omega}^{2}\right){\gamma}^{2}}+\cr
&&  \frac{\frac{1}{4}{\epsilon}^{2}{\Omega}^{2} -6\alpha \beta \gamma{\epsilon}^{2}{\Omega}^{2}
  -\alpha \beta {\epsilon}^{2}{\Omega}^{2}+
  {\epsilon}^{2} {\alpha}^{2} {\beta}^{2}{\Omega}^{2}}
  {\frac{9}{16}\left(9{\gamma}^{2}+{\epsilon}^{2}{\Omega}^{2}\right){\gamma}^{2}}, \label{eq29}\\
  P_{3}&=&\frac{4\left(1-{\Omega}^{2}\right)\left(\frac{9}{2}\gamma
 -3\gamma{\Omega}^{2}-2{\beta}^{2}\right)}
  {\frac{9}{16}\left(9{\gamma}^{2}+{\epsilon}^{2}{\Omega}^{2}\right){\gamma}^{2}}+\cr
&&+\frac{6{\epsilon}^{2}{\Omega}^{2}+6{\epsilon}^{2}{\Omega}^{2}\gamma
 -4\alpha \beta{\epsilon}^{2}{\Omega}^{2}-9{\gamma}^{2}F^{2}}
 {\frac{9}{16}\left(9{\gamma}^{2}+{\epsilon}^{2}{\Omega}^{2}\right){\gamma}^{2}}, \label{eq30}\\
 P_{4}&=&\frac{4\left(1-{\Omega}^{2}\right)^{2}-6\gamma F^{2}}
 {\frac{9}{16}\left(9{\gamma}^{2}+{\epsilon}^{2}{\Omega}^{2}\right){\gamma}^{2}}.\label{eq31}
\end{eqnarray}

\begin{figure}[htbp]
\begin{center}
 \includegraphics[width=12cm,  height=6cm]{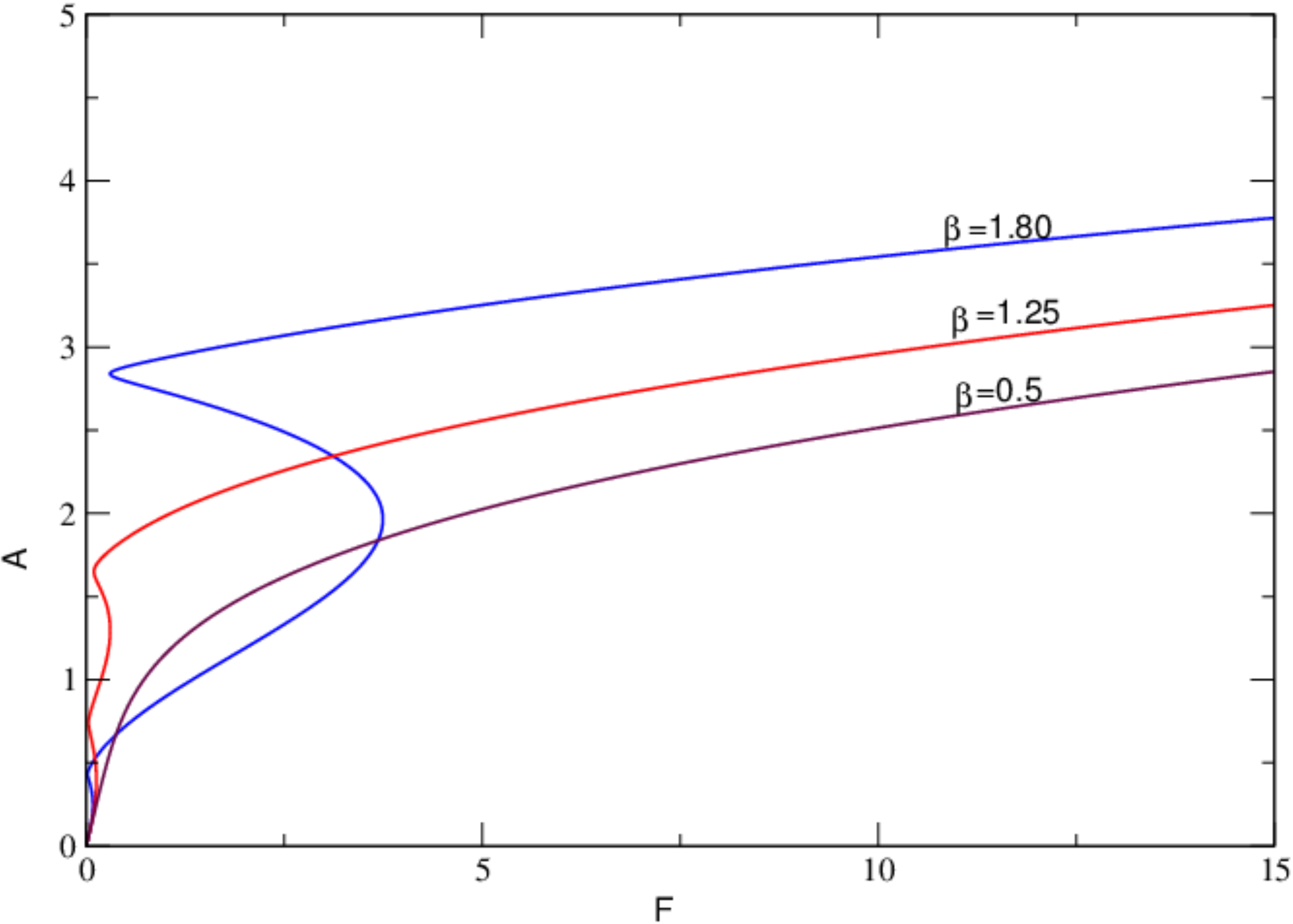}
\end{center}
\caption{Effects of $\beta $ on the amplitude-response curves with 
$ \Omega=0.7; \gamma=0; \epsilon=0.05$ and $\alpha=0$.}
\label{fig:1}
\end{figure}

\begin{figure}[htbp]
\begin{center}
 \includegraphics[width=12cm,  height=6cm]{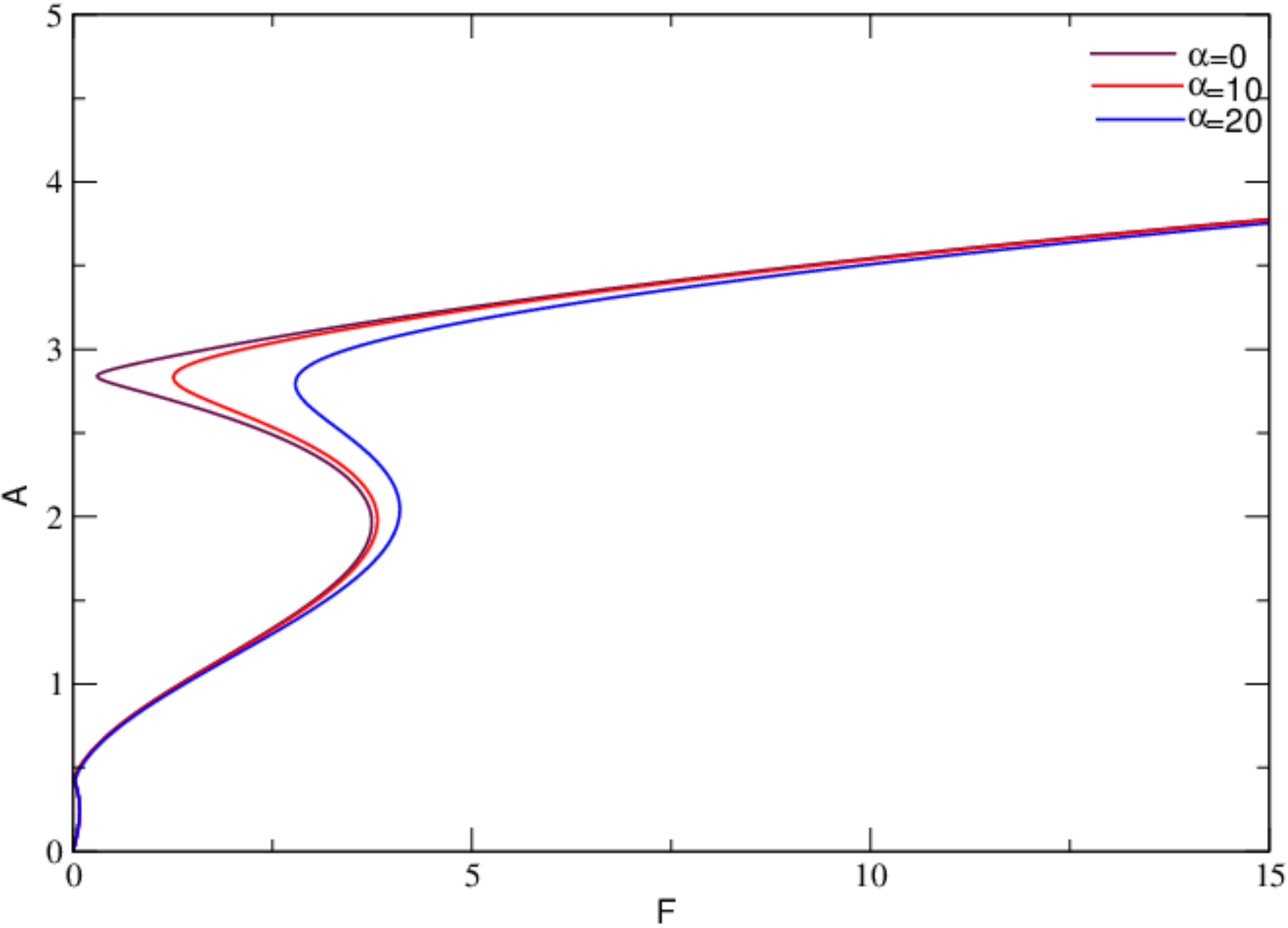}
\end{center}
\caption{Effects of $\alpha $ on the amplitude-response curves with $ \Omega=0.7; \gamma=0; \epsilon=0.05$ and $\beta=1.85$
.}
\label{fig:2}
\end{figure}

\begin{figure}[htbp]
\begin{center}
 \includegraphics[width=12cm,  height=6cm]{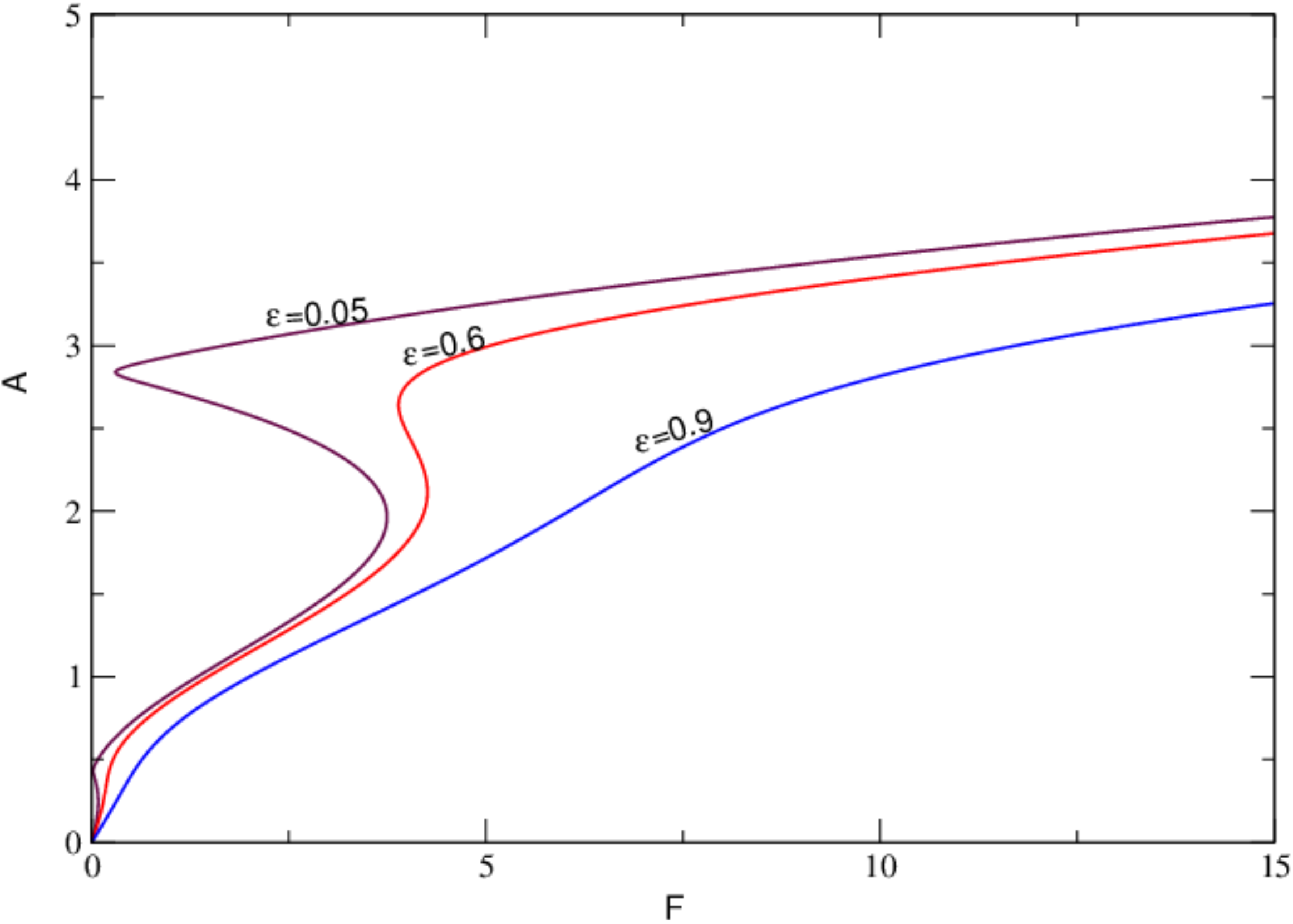}
\end{center}
\caption{Effects of $\epsilon$ on the amplitude-response curves with $ \Omega=0.7; \gamma=0; \beta=1.85$ and $\alpha=0$.}
\label{fig:3}
\end{figure}

We investigate the effects of the quadratic and cubic terms on the behavior of the amplitude of plasma oscillations. The obtained
results are repoted in Figs. \ref{fig:1}, \ref{fig:2} and \ref{fig:3} where the hysteresis and jump phenomena are found.
 It should be stressed that the hysteresis and jump phenomena known
to be function of the cubic nonlinear coefficient can also be triggered or quenched through the
dissipative coefficient $\epsilon$ (see Figs. \ref{fig:3}) and as well as via the nonlinear quadratic parameter $\beta$ and $\alpha$ (see
Figs. \ref{fig:1}, \ref{fig:2} ). Through (\ref{eq27}), the behavior of
 the amplitude of plasma oscillations is investigated when the external frequency $\Omega$
varies. Thereby, the observed resonant state obtained
for a set of parameters can be destroyed according to the value taken by the amplitude of the
external force. During the hysteresis and jump phenomena processes, for any value of the frequency $\Omega$
and the external force $F$ respectively, three different amplitudes of oscillations are obtained among
which, two are stable and one is unstable.

\newpage

\section{Resonant states}
We investigate the differents resonances with the Multiple time scales Method $(MSM)$.

In such a situation,  an approximate solution is generally sought as follows:
\begin{eqnarray}
 x(\epsilon,  t)&=&x_{0}(T_{0},  T_{1} )+ \epsilon x_{1}(T_{0},  T_{1} )+ \dots \label{eq32}
\end{eqnarray}

With $T_{n}={\epsilon}^{n}t$
The derivatives operators can now be rewritten as follows:
\begin{eqnarray}
 \frac{d}{dt}&=&D_{0} +\epsilon D_{1}+\dots \label{eq33}\\
 \frac{d^{2}}{{dt}^{2}}&=&D_{0}^{2}+2\epsilon D_{0}D_{1} +\dots\label{eq34} 
\end{eqnarray}

where ${D^{m}_{n}}=\frac{\partial^{m}}{\partial{T_{n}^{m}}}$
\subsection{Primary resonant state}
In this state,  we put that 
$F=\epsilon F$,  $\beta=\epsilon \beta$,  $\gamma =\epsilon\gamma$.
 The closeness between both internal and external frequencies is given by 
 $\Omega = 1+\epsilon \sigma$. Where $\sigma$ is the detuning parameter. Inserting (\ref{eq34}) 
 and (\ref{eq35})
 into (\ref{eq20}) we obtain:
 \begin{eqnarray}
   \left(D_{0}^{2}+2\epsilon D_{0}D_{1}\right)(x_{0}+ \epsilon x_{1})+\nonumber\\
   +\epsilon \left[ 1 +{\left(x_{0}+ \epsilon x_{1}\right)}^{2} \right]
   \left(D_{0} +\epsilon D_{1}\right)(x_{0}+\epsilon x_{1} )+\nonumber\\
  +x_{0}+ \epsilon x_{1} + \epsilon \alpha (x_{0}+\epsilon x_{1} )\left(D_{0} +\epsilon D_{1}\right)
   (x_{0}+{\epsilon} x_{1} ) +\nonumber\\
   +{\beta}(x_{0}+\epsilon x_{1} )^{2}+
  \gamma (x_{0}+\epsilon x_{1} )^{3}
 -{\epsilon} F\cos{\Omega t}&=&0; \label{eq35}
 \end{eqnarray}

 Equating the the coefficients of like powers of $\epsilon$ after some algebraic 
 manipulations,  we obtain:
 
 In order ${\epsilon}^{0}$, 
 \begin{eqnarray}
  D^{2}_{0}x_{0}+x_{0}&=&0\label{eq36} 
  \end{eqnarray}
In order ${\epsilon}^{1}$, 
\begin{eqnarray}
 D^{2}_{0}x_{1}+x_{1}&=&F\cos{\Omega}t-2D_{1}D_{0}x_ {0}- \left( 1+x_{0}^{2}\right)D_{0}x_{0}-
 {\alpha}x_{0}D_{0}x_{0}\cr
 &&-{\beta}x_{0}^{2}-
 {\gamma}x_{0}^{3}.\label{eq37}
\end{eqnarray}
The general solution of (\ref{eq36}) is 
\begin{eqnarray}
 x_{0}&=&A(T_{1})\exp(jT_{0})+CC,  \label{eq38}
\end{eqnarray}
where $CC$ represents the complex conjugate of the previous terms.
$A(T_{1} )$ is a complex function to be determined from solvability or secular conditions 
of (\ref{eq37}). Thus,  substituting the solution $x_{0}$ in (\ref{eq38})
leads us to the following secular criterion

\begin{eqnarray}
 2j{A'}+ j\left(1+{|A|^{2}}\right)A+ 3\gamma{|A|^{2}}A-\frac{F}{2}\exp{(j{\sigma T_{1}})}&=&0 
 \label{eq39}
\end{eqnarray}
In polar coordinates,  the solution of  (\ref{eq39}) is
\begin{eqnarray}
 A&=&\frac{1}{2}a(T_{1})\exp{\left[j\theta(T_{1})\right]}\label{eq40}
\end{eqnarray}

where $a$ and $\theta$ are real quantities and stand respectively for the amplitude and phase
of oscillations. After injecting (\ref{eq40}) into (\ref{eq39}),  we separate real and
imaginary terms and obtain the following coupled flow for the amplitude and phase:

\begin{eqnarray}
{a'}= -\frac{a}{2}-\frac{a^{3}}{8} +\frac{F}{2}\sin{\Phi}
\label{eq41}    \\              
a{\Phi'}=a \sigma-\frac{3 \gamma  a^{3}}{8}-\frac{F}{2}\cos{\Phi}.\label{eq42}   
\end{eqnarray}
where the prime denotes the derivative with respect to $T_{1}$ and 
$\Phi = {\sigma} {T_{1}} - {\theta}$. For the steady-state
conditions $({a'} = {\Phi'} = 0)$,  the following nonlinear algebraic equation is obtained :

\begin{eqnarray}
 \left(\frac{9{\gamma}^{2}+1}{64}\right) {a_{0}^{6}}+
 \left(\frac{1-6\gamma \sigma}{8}\right) {a_{0}^{4}}+
 \left(\frac{1+4\sigma^{2}}{4}\right) {a_{0}^{2}}-\frac{F^{2}}{4}&=&0\label{eq43}
\end{eqnarray}

where ${a_{0}}$ and ${\Phi_{0}}$ are respectively the values of 
$a$ and $\Phi$ in the steady-state. 
 Eq.(\ref{eq43}) is the equation of primary resonance flow.
 Now,  we study the  stability of the precess,  we assume that each equilibrium state 
 is submitted to a small pertubation as follows
\begin{eqnarray}
 a&=&a_{0}+a_{1}\label{eq44}\\
 \Phi&=&\Phi_{0}+\Phi_{1}\label{eq45}
\end{eqnarray}

where $a_{1}$ and $\Phi_{1}$ are slight variations. 
Inserting  the equations (\ref{eq44}) and 
(\ref{eq45}) into (\ref{eq41}) and (\ref{eq42})
  and canceling nonlinear
terms enable us to obtain

\begin{eqnarray}
{a'}_{1}= -\frac{1}{2}\left(1+\frac{3a_{0}^{2}}{4}\right)a_{1}-
a_{0}\left(\sigma-\frac{3a_{0}^{2}\gamma}{8}\right){\Phi}_{1}
\label{eq46}    \\              
{\Phi'}_{1}=\frac{1}{a_{0}}\left(\sigma-\frac{9a_{0}^{2}\gamma}{8}\right)a_{1}
-\frac{1}{2}\left(1+\frac{a_{0}^{2}}{4}\right){\Phi}_{1}.\label{eq47}   
\end{eqnarray}
The stability process depends on the sign of eigenvalues $\varUpsilon$ 
of the equations (\ref{eq46}) and (\ref{eq47}) which are given through
the following characteristic equation
\begin{eqnarray}
 \varUpsilon^{2}+2Q\varUpsilon+R&=&0, \label{eq48}
\end{eqnarray}
where
$Q=\frac{1}{4}\left( a_{0}^{2}+2\right)$,  
$R=\frac{1}{4}\left(1+\frac{3a_{0}^{2}}{4}\right)\left(1+\frac{a_{0}^{2}}{4}\right)-
\left(\sigma-\frac{3a_{0}^{2}\gamma}{8}\right)\left(\sigma-\frac{9a_{0}^{2}\gamma}{8}\right)$.

\begin{figure}[htbp]
\begin{center}
 \includegraphics[width=12cm,  height=6cm]{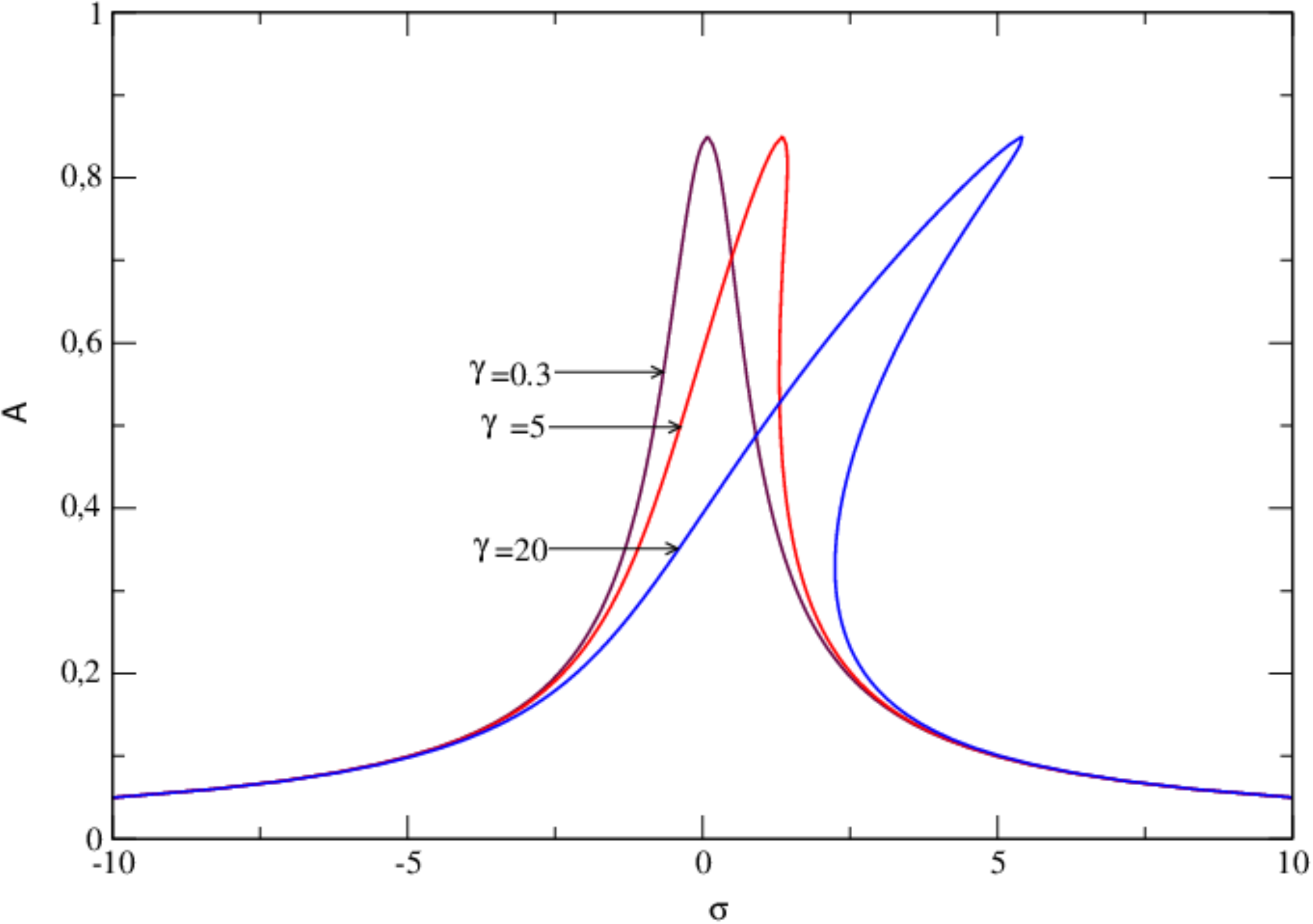}
\end{center}
\caption{Effects of $\gamma$ on the frequency-response curves of the primary resonance with the parameters for F = 1.0.}
\label{fig:4}
\end{figure}

Since$ Q > 0$,  the steady-state solutions are stable if $R > 0$ and unstable otherwise.
Fig. \ref{fig:4} displays
the amplitudes response curves obtained from (\ref{eq43}) for 
different values of the parameter $\gamma$ and
one can observe that as it increases,  the model goes from resonance to a hysteresis state.

\newpage
\subsection{Superharmonic and subharmonic oscillations}
When the amplitude of the sinusoidal external force is large,  other type of oscillations can be
displayed by the model,  namely the superharmonic and the subharmonic oscillatory states. It is
now assumed that $F = \epsilon^{0}F$ and therefore,  one obtains the following equations at 
different order of $\epsilon$.
In order ${\epsilon}^{0}$, 
 \begin{eqnarray}
  D^{2}_{0}x_{0}+x_{0}&=&F_{0}\cos{\Omega T_{0}}\label{eq49} 
  \end{eqnarray}
In order ${\epsilon}^{1}$, 
\begin{eqnarray}
 D^{2}_{0}x_{1}+x_{1}&=&-2D_{1}D_{0}x_ {0}- \left( 1+x_{0}^{2}\right)D_{0}x_{0}-
 {\alpha}x_{0}D_{0}x_{0}\cr
 &&-{\beta}x_{0}^{2}-
 {\gamma}x_{0}^{3}.\label{eq50}
\end{eqnarray}

The general solution of Eq. (\ref{eq50}) is
 \begin{eqnarray}
  x_{0}&=&A(T_{1})e^{jT_{0}}+{\Lambda}e^{j{\Omega}T_{0}},  \label{eq51}
 \end{eqnarray}
with $\Lambda=\frac{F}{2(1-{\Omega}^{2})}$

Substituting the general solution $x_{0}$ into Eq. (\ref{eq51}),  after some
 algebraic manipulations,  we obtain
\begin{eqnarray}
  D^{2}_{0}x_{1}+x_{1}&=&\left[-2j {A'} -\left(j \left(1+|A|^{2} + 2{\Lambda}^{2}\right) +
  3{\gamma}|A|^{2}+6{\gamma}{\Lambda}^{2}\right)\right] e^{jT_{0}}+\cr
  &&-\left[ j\Omega \left(1+{\Lambda}^{2}+2|A|^{2}\right)+ 3\gamma |\Lambda|^{2}+
  6\gamma |A|^{2}\right]\Lambda e^{j\Omega T_{0}} +\cr
  && -\left(j+\gamma\right)A^{3}e^{3j T_{0}}-
  \left[j\left( 1+2\Omega\right)+3\gamma\right]A{\Lambda}^{2} e^{j(1+2\Omega)T_{0}}+\cr
  && -\left[j\left(2+\Omega \right) + 3\gamma \right] A^{2}\Lambda e^{j(2+\Omega)T_{0}}+\cr
  &&-\left[j\left(2-\Omega\right) +3\gamma\right]A^{2}{\Lambda}e^{j(2-\Omega)T_{0}}-\left(j\Omega
 + \gamma\right){\Lambda}^{3}e^{j3{\Omega}T_{0}}+\cr
 &&-\left[j\left(1-2\Omega\right) +3\gamma\right]A{\Lambda}^{2}e^{j(1-2\Omega)T_{0}}
 -\left(j\alpha+ \beta\right){A}^{2}e^{2jT_{0}}+\cr
 &&-\left(j\alpha \Omega +\beta\right){\Lambda}^{2}e^{2j\Omega T_{0}}-
 \left[j\alpha\left(1+\Omega\right) +2\beta\right] A{\Lambda} e^{j(1+\Omega)T_{0}}\cr
 &&-\left[j\alpha\left(1-\Omega\right) +2\beta\right] A{\Lambda} e^{j(1-\Omega)T_{0}}
 -2\beta\left(|A|^{2} +\Lambda^{2}\right) +\cr
 &&+j\alpha\left(|A|^{2}+\Omega \Lambda^{2}\right) +CC \label{eq52}
\end{eqnarray}

From Eq.(\ref{eq52}),  it comes that superharmonic and subharmonic states can be found from the
quadratic and cubic nonlinearities. The cases of superhamonic oscillations we consider are
$3\Omega = 1 + \epsilon \sigma$ and $2\Omega = 1 + \epsilon \sigma$,  while the subharmonic 
oscillations to be treated are $\Omega = 3 + \epsilon \sigma$
and $\Omega = 2 + \epsilon \sigma$.

For the first superhamonic states $3\Omega = 1 + \epsilon \sigma$,  equating resonant terms at $0$
 from Eq.(\ref{eq52}),  we obtain:

\begin{eqnarray}
 -2j {A'}-j\left( 1 +|A|^{2} + 2{\Lambda}^{2}\right)A -3\gamma |A|^{2}A-6\gamma{\Lambda}^{2}A
 \nonumber\\
 -\left(j\Omega +\gamma\right){\Lambda}^{3}e^{j\epsilon \sigma T_{0}}&=&0\label{eq53}
\end{eqnarray}

  Using (\ref{eq40}) and after some algebraic manipulations,  we rewritte (\ref{eq53}) as follows 
 
 \begin{eqnarray}
  {a'}&=&-\frac{a}{2}-\frac{a^3}{8}-a{\Lambda}^{2}-
 { \Lambda}^{3}\left({\Omega} \cos{\Phi}+{\gamma }\sin{\Phi}\right), \label{eq54}\\
  a{\Phi'}&=& a\sigma -\frac{3\gamma {a}^3}{8}- 3\gamma {\Lambda}^{2}a +
  {\Lambda}^{3}\left(\gamma \cos{\Phi-\Omega\sin{\Phi}}\right)\label{eq55}
 \end{eqnarray}

 The amplitude of oscillations of this superharmonic states  is governed by the 
 following nonlinear algebraic equation

\begin{eqnarray}
 \left(\frac{a_{0}}{2}+\frac{a_{0}^3}{8}+a_{0}{\Lambda}^{2}\right)^{2} +
 \left(a_{0}\sigma -\frac{3\gamma {a}_{0}^3}{8}- 3\gamma {\Lambda}^{2}a_{0}\right)^{2}-
 \nonumber\\
 \left({\Omega}^{2}+{\gamma}^{2}\right)\Lambda^{2}&=&0. \label{eq56}
\end{eqnarray}
After some algebraic manipulations,  Eq.(\ref{eq56}) can be rewritten as follows
\begin{eqnarray}
  \left(\frac{9{\gamma}^{2}+1}{64}\right)a_{0}^{6}+
 \frac{1}{4}\left[\frac{1}{2}+{\Lambda}^{2}-3\gamma 
 \left( \sigma-3\gamma \Lambda^{2}\right)\right]a_{0}^{4}
 &=&\nonumber\\
 -\left[\left(\frac{1}{2}+{\Lambda}^{2}\right)^{2}+
 \left(\sigma- 3\gamma {\Lambda}^{2} \right)^{2}\right]a_{0}^{2}+ 
\left(\gamma^{2}+{\Omega}^{2}\right){\Lambda}^{6}\label{eq57}
\end{eqnarray}
and they are stable if
\begin{eqnarray}
 \frac{1}{4}\left( 1+2\Lambda^{2}+ \frac{3}{4}a_{0}^{2}\right) \left(1+2\Lambda^{2}+
 \frac{1}{4}a_{0}^{2}\right) +\nonumber\\
 \left(\sigma-\frac{9}{8}\gamma a_{0}^{2}-3\gamma \Lambda^{2}\right)
 \left(\sigma-3\gamma \Lambda^{2}-\frac{3}{8}\gamma a_{0}\right)>0\label{eq58}
\end{eqnarray}

\begin{figure}[htbp]
\begin{center}
 \includegraphics[width=12cm,  height=6cm]{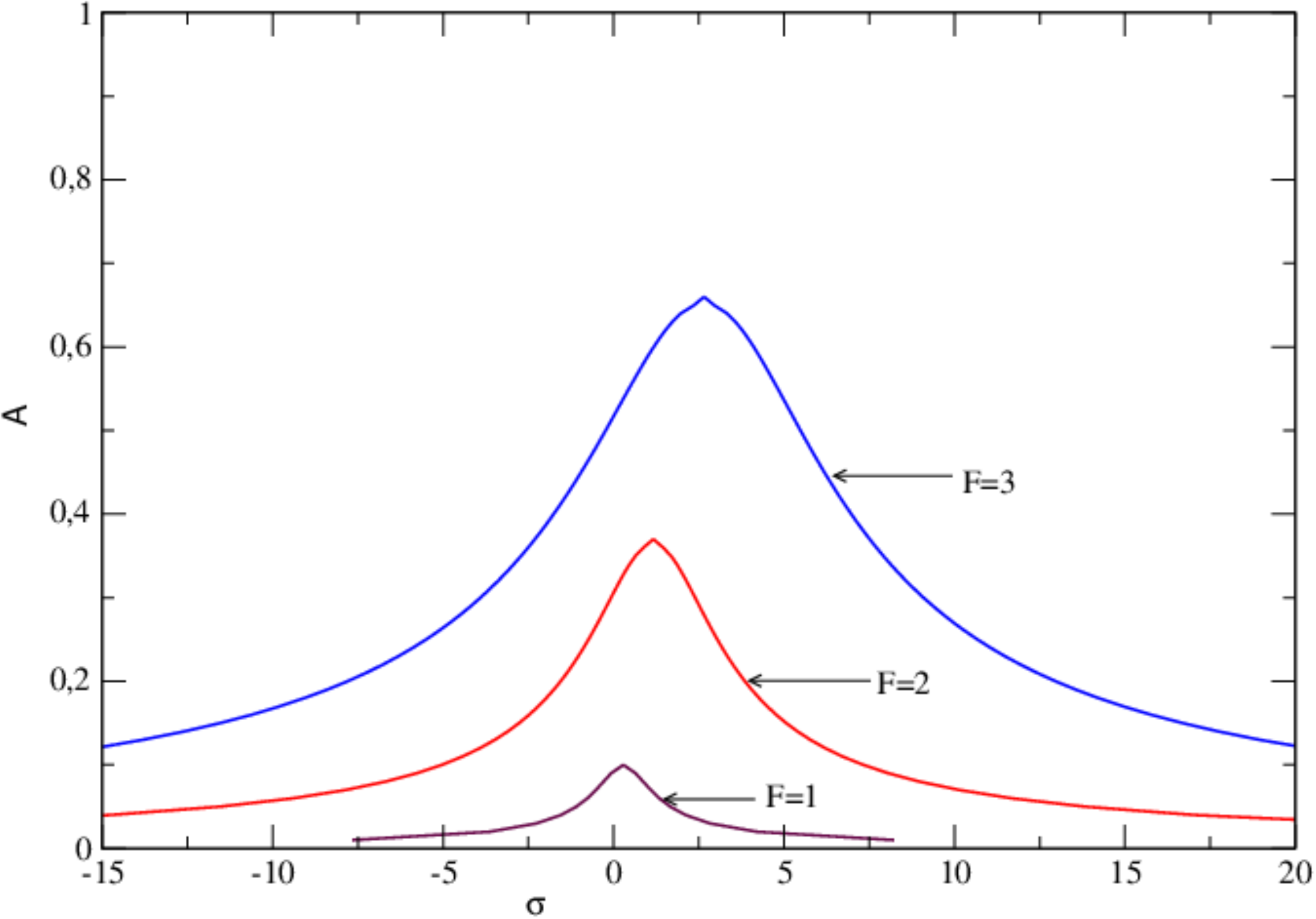}
\end{center}
\caption{Effects of $F$ on the frequency-response curves of the order-three superharmonic resonance with $\gamma=0.3$}
\label{fig:5}
\end{figure}

\begin{figure}[htbp]
\begin{center}
 \includegraphics[width=12cm,  height=6cm]{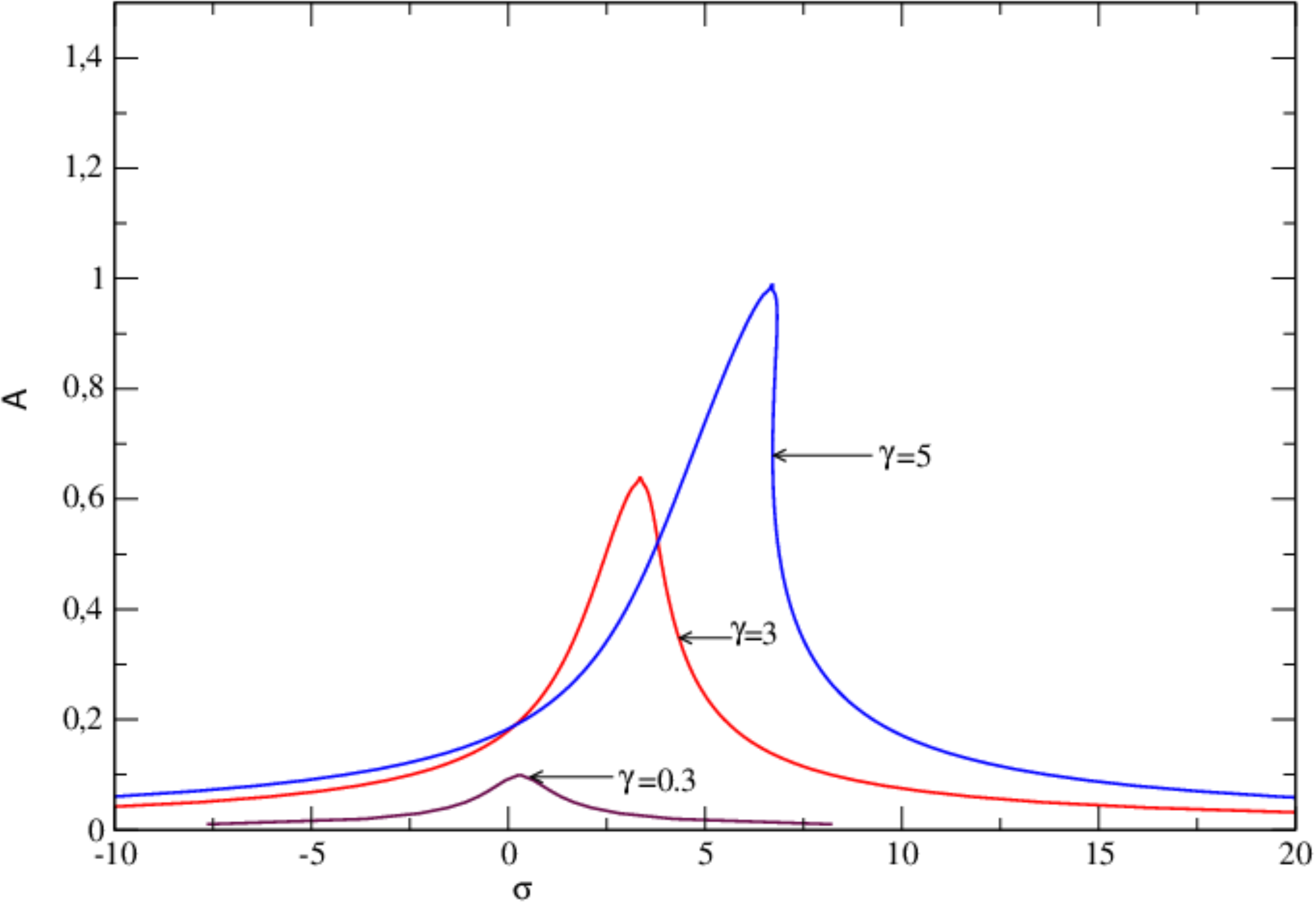}
\end{center}
\caption{Effects of $\gamma$ on the frequency-response curves of the order-three superharmonic resonance with $F=1$.}
\label{fig:6}
\end{figure}

Fig. \ref{fig:5} presents the frequency response curves of the superharmonic resonance as a function of
$\sigma$ for different values of the external force. As the intensity of the external force increases,  the
resonance behavior observed is destroyed. The effects of the parameter $\gamma$ on such superharmonic
oscillations are also investigated and results are reported in Fig.\ref{fig:6},  showing the appearance of the
hysteresis phenomenon when the nonlinear cubic parameter $\gamma$ is increasing.

\begin{figure}[htbp]
\begin{center}
 \includegraphics[width=12cm,  height=6cm]{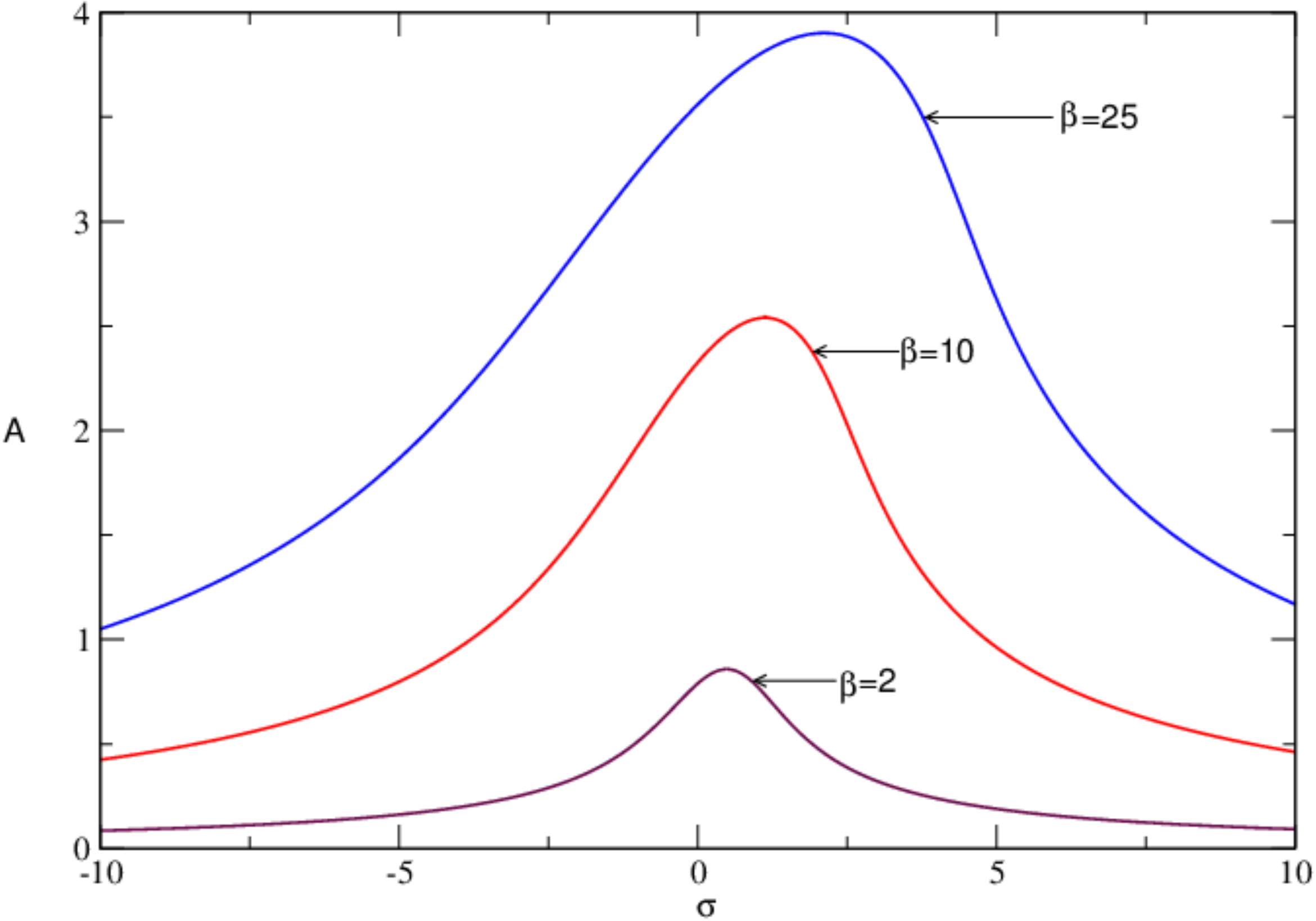}
\end{center}
\caption{Effects of $\beta$ on the frequency-response curves of the order-two superharmonic resonance with $F=1$.}
\label{fig:61}
\end{figure}

\begin{figure}[htbp]
\begin{center}
 \includegraphics[width=12cm,  height=6cm]{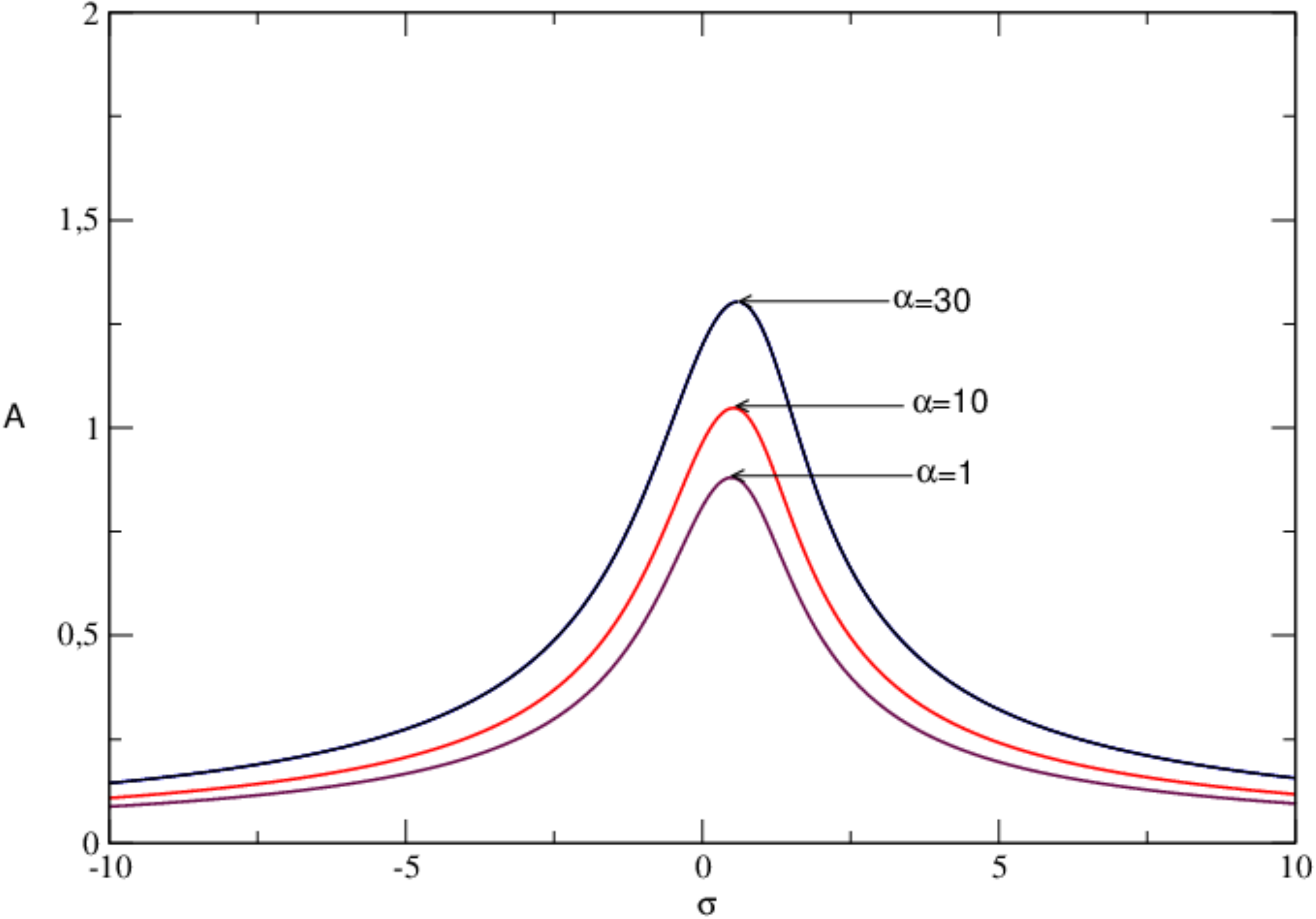}
\end{center}
\caption{Effects of $\alpha$ on the frequency-response curves of the order-two superharmonic resonance with $F=1$ and $\beta=2$.}
\label{fig:62}
\end{figure}

 On the other hand,  the second superharmonic states $2\Omega = 1 + \epsilon \sigma$,  
 inserting this condition into Eq.(\ref{eq52}) and equating the secular terms to $0$,  we obtain:
 \begin{eqnarray}
  -2j {A'}-j\left( 1 +|A|^{2} + 2{\Lambda}^{2}\right)A +3\gamma |A|^{2}A+6\gamma{\Lambda}^{2}A
 \nonumber\\
 -\left(j\alpha \Omega +\beta\right){\Lambda}^{2}e^{j\epsilon \sigma T_{0}}&=&0\label{eq59}
 \end{eqnarray}
After some algebraic manipulations as in the first superhamonic resonance,  we obtain:
 \begin{eqnarray}
 \left(\frac{9{\gamma}^2+1}{64}\right)a_{0}^{6}+
 \frac{1}{4}\left[\frac{1}{2}+{\Lambda}^2-3\gamma 
 \left( \sigma-3\gamma \Lambda^{2}\right)\right]a_{0}^{4}
 &=&-\nonumber\\
 +\left[\left(\frac{1}{2}+{\Lambda}^{2}\right)^{2}+
 \left(\sigma- 3\gamma {\Lambda}^{2} \right)^{2}\right]a_{0}^{2}+ 
\left(\alpha^{2} \omega^{2}+{\beta}^{2}\right){\Lambda}^{6}, \label{eq60} 
 \end{eqnarray}

 and the stability criteria  is the one defined through inequality (\ref{eq58}). 
 In this cases also,  the influence  of the quadratic parameter on such oscillations has been checked
 (see Figs \ref{fig:61} and \ref{fig:62} ). In such a state, we noticed that  the peak values of this superharmonic resonance
 are increased  progressively when increasing $\beta$ or $\alpha$. 
 
 The first subharmonic oscillations $(\Omega=3+\epsilon \sigma)$ are fund.
 Inserting this condition into Eq.(\ref{eq52}) and equating the secular
  terms to $0$,  we obtain:

 \begin{eqnarray}
  -2j {A'}-\left[j\left(1+ |A|^{2}+ 2\Lambda^{2}\right)+3{\gamma}|A|^{2}+6\gamma \Lambda^{2}\right]A
 \nonumber\\
 - \left(j\left(2-\Omega\right)+3\gamma\right){\Lambda} {\bar{A}}^{2} e^{j\sigma T_{0}}&=&0
 \label{eq61}
 \end{eqnarray}

 With (\ref{eq40}),  Eq.(\ref{eq61}) gives  after some algebraic manipulations and separating 
 real and imaginary terms, 
 \begin{eqnarray}
  {a'}&=&-\frac{a}{2}-\frac{a^3}{8}-a{\Lambda}^{2}-
\frac{1}{4} { \Lambda} a^{2}\left[(2-{\Omega}) \cos{\Phi}+3{\gamma }\sin{\Phi}\right], \label{eq62}\\
  a{\Phi'}&=&\frac{1}{3} a\sigma -\frac{3\gamma {a}^3}{8}- 3\gamma {\Lambda}^{2}a +
 \frac{1}{4} {\Lambda} a^{2}\left[3\gamma \cos{\Phi-(2-{\Omega})\sin{\Phi}}\right]\label{eq63}
 \end{eqnarray}

  Study-states oscillations conditions implies
  \begin{eqnarray}
   \left(\frac{a_{0}}{2}+\frac{a_{0}^3}{8}+a_{0}{\Lambda}^{2}\right)^{2}+
   \left(\frac{1}{3} a_{0}\sigma -\frac{3\gamma {a}_{0}^3}{8}- 
   3\gamma {\Lambda}^{2}a_{0} \right)^{2} \nonumber\\
   -\frac{1}{16}{\Lambda}^{2}{a_{0}}^{4}\left[(2-\Omega)^{2}+9{\gamma}^{2}\right]&=&0\label{eq64}
  \end{eqnarray}

 Finally the amplitude of subharmonic oscillatory states are given by following nonlinear algebraic
 equation:
 \begin{eqnarray}
  \left(\frac{9{\gamma}^{2}+1}{64}\right)a_{0}^{4} + \nonumber\\ 
  +\left[\frac{1}{8} + \frac{\Lambda^{2}}{4}
  -\frac{3}{4}\gamma \left(\frac{\sigma}{3}-3\gamma \Lambda^{2}\right) -\frac{1}{16}\Lambda^{2}
  \left(9\gamma^{2}+\left(2-\Omega \right)^{2} \right)\right] a_{0}^{2}+ \nonumber\\
 +\left[\left(\frac{1}{2} +\Lambda^{2}\right)^{2}+
 \left(\frac{\sigma}{3}-3\gamma \Lambda^{2}\right)^{2}\right]&=&0
 \label{eq65}
 \end{eqnarray}

 and the stability is guaranted only if 
 
 \begin{eqnarray}
  \left( \frac{1}{2} + \Lambda^{2} +\frac{1}{4}a_{0}^{2}\right)
  \left( \frac{1}{2} + \Lambda^{2} +\frac{1}{8}a_{0}^{2}\right)+ \nonumber\\-\left(\frac{\sigma}{9}
  -\gamma \Lambda^{2} -\frac{1}{8}a_{0}^{2}\right)\left(\sigma +
  \frac{9}{8}\gamma a_{0}^{2}-9\gamma\Lambda^{2}\right)<0\label{eq66}
 \end{eqnarray}

 Now,  we consider the second subharmonic oscillatory states: $\Omega= 2 +\epsilon\sigma$. This condition
 implies althrough Eq.(\ref{eq52}) the secular term which equating $0$  give
 
\begin{eqnarray}
  -2j {A'}-\left[j\left(1+ |A|^{2}+ 2\Lambda^{2}\right)+3{\gamma}|A|^{2}+6\gamma \Lambda^{2}\right]A
 \nonumber\\
 - \left(j\left(1-\Omega\right)\alpha+
 2\beta\right] {\bar{A}}\Lambda e^{j\sigma T_{1}}&=&0
 \label{eq67}
 \end{eqnarray}
 We obtain that the second subharmonic oscillatory states motions are governed by the equation
 
 \begin{eqnarray}
  \left(\frac{9{\gamma}^{2}+1}{64}\right)a_{0}^{4} +\nonumber \\ \left[\frac{1}{8} + \frac{\Lambda^{2}}{4}
  -\frac{3}{4}\gamma \left(\frac{\sigma}{2}-3\gamma \Lambda^{2}\right) -\frac{1}{16}\Lambda^{2}
  \left(9\gamma^{2}+\left(2-\Omega \right)^{2} \right)\right] a_{0}^{2}+\nonumber\\
 \left[\left(\frac{1}{2} +\Lambda^{2}\right)^{2}+
 \left(\frac{\sigma}{2}-3\gamma \Lambda^{2}\right)^{2} 
 -\left(1-\Omega\right)^{2}\alpha^{2} \Lambda^{2}-4\beta^{2} \Lambda^{2}\right]&=&0\nonumber\\
 \label{eq68}
 \end{eqnarray}

 They are stable if the following criterion is fulfilled
 \begin{eqnarray}
   \left( \frac{1}{2} +2 \Lambda^{2} +\frac{1}{4}a_{0}^{2}\right)a_{0}^{2}+
  6\left[\left(\frac{\sigma}{2}-3\gamma \Lambda^{2}\right)a_{0}^{2}
  -\frac{3}{8}\gamma a_{0}^{4}\right]>0 \label{eq69}
 \end{eqnarray}

\begin{figure}[htbp]
\begin{center}
 \includegraphics[width=12cm,  height=6cm]{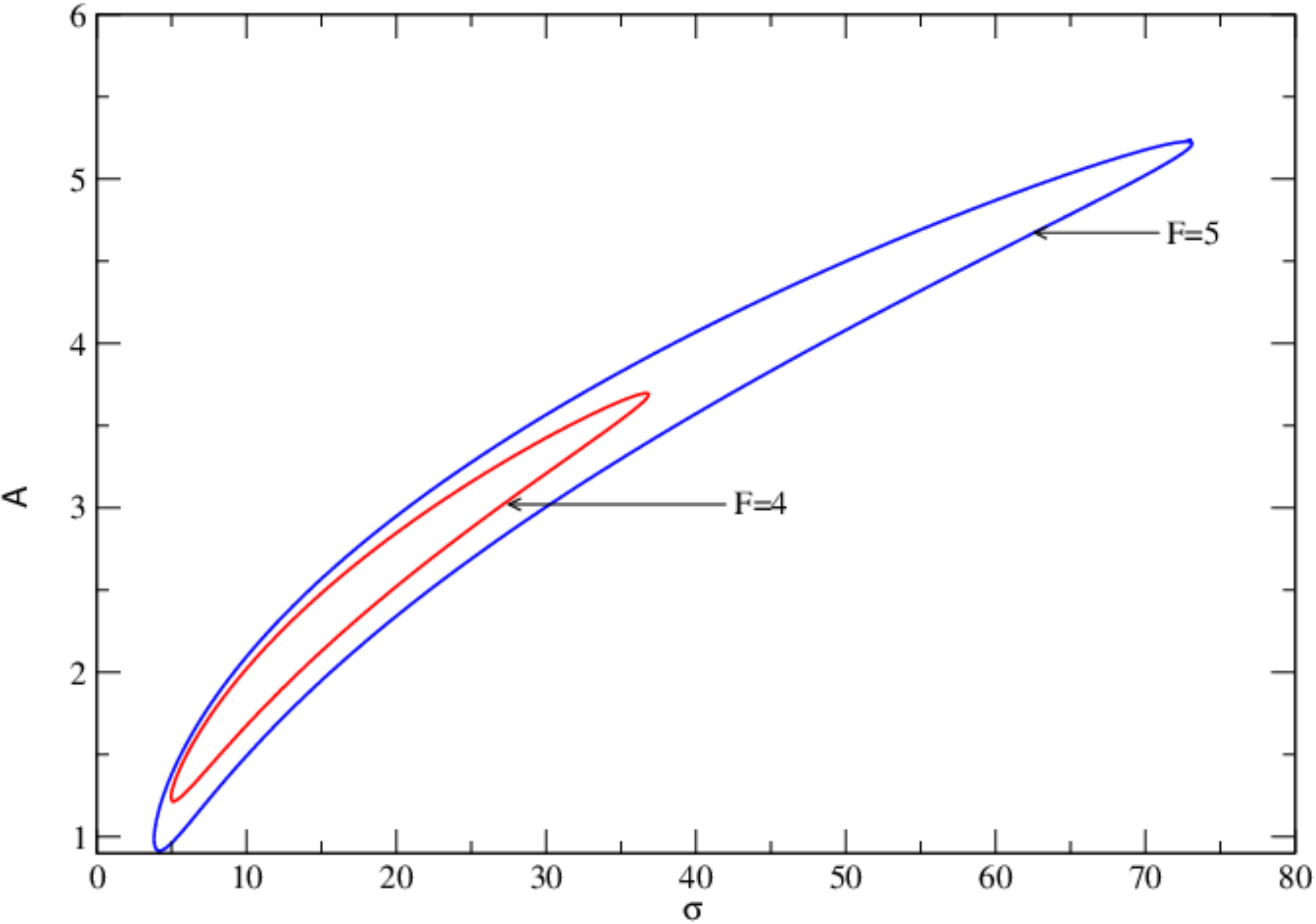}
\end{center}
\caption{Effects of $F$ on the frequency-response curves of the order-three subharmonic resonance with $\gamma = 2.30$.}
\label{fig:7}
\end{figure}

\begin{figure}[htbp]
\begin{center}
 \includegraphics[width=12cm,  height=6cm]{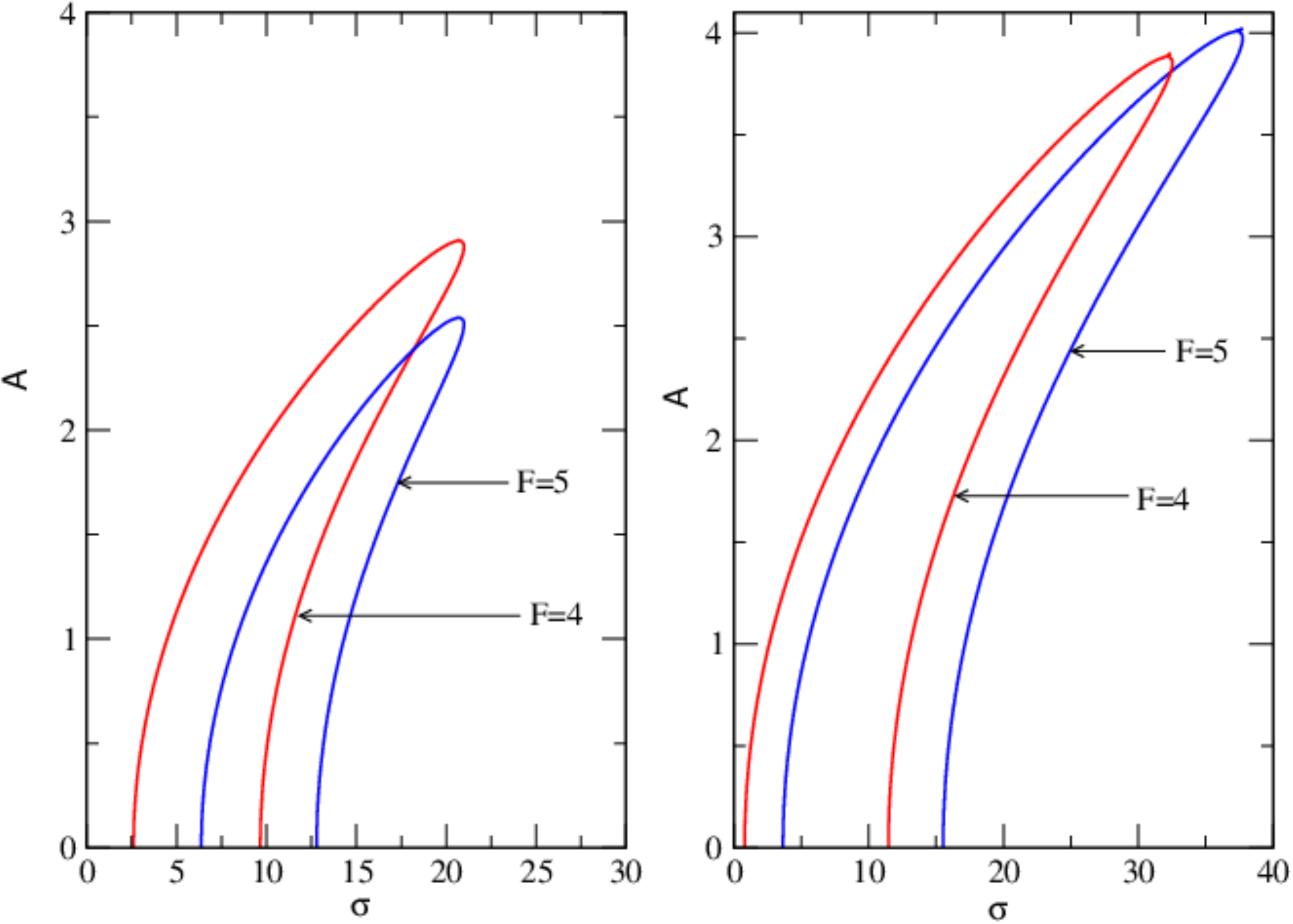}
 
\end{center}
\caption{Effects of $F$ on the frequency-response curves of the first subharmonic 
resonance with $\gamma=2.30; \beta=1$;left $\alpha=0$ and right $\alpha=1$.}
\label{fig:8}
\end{figure}

The frequency response curves of both types of subharmonic oscillations are plotted in Figs.\ref{fig:7} and \ref{fig:8} and the regions
 where such behaviors occur are obtained. From these pictures,  it comes that
the range of frequency where a response can be obtained is more important in the first subharmonic
state than in the other cases.


\newpage

 \section{Bifurcation and Chaotic behavior}
\begin{figure}[htbp]
\begin{center}
 \includegraphics[width=12cm,  height=10cm]{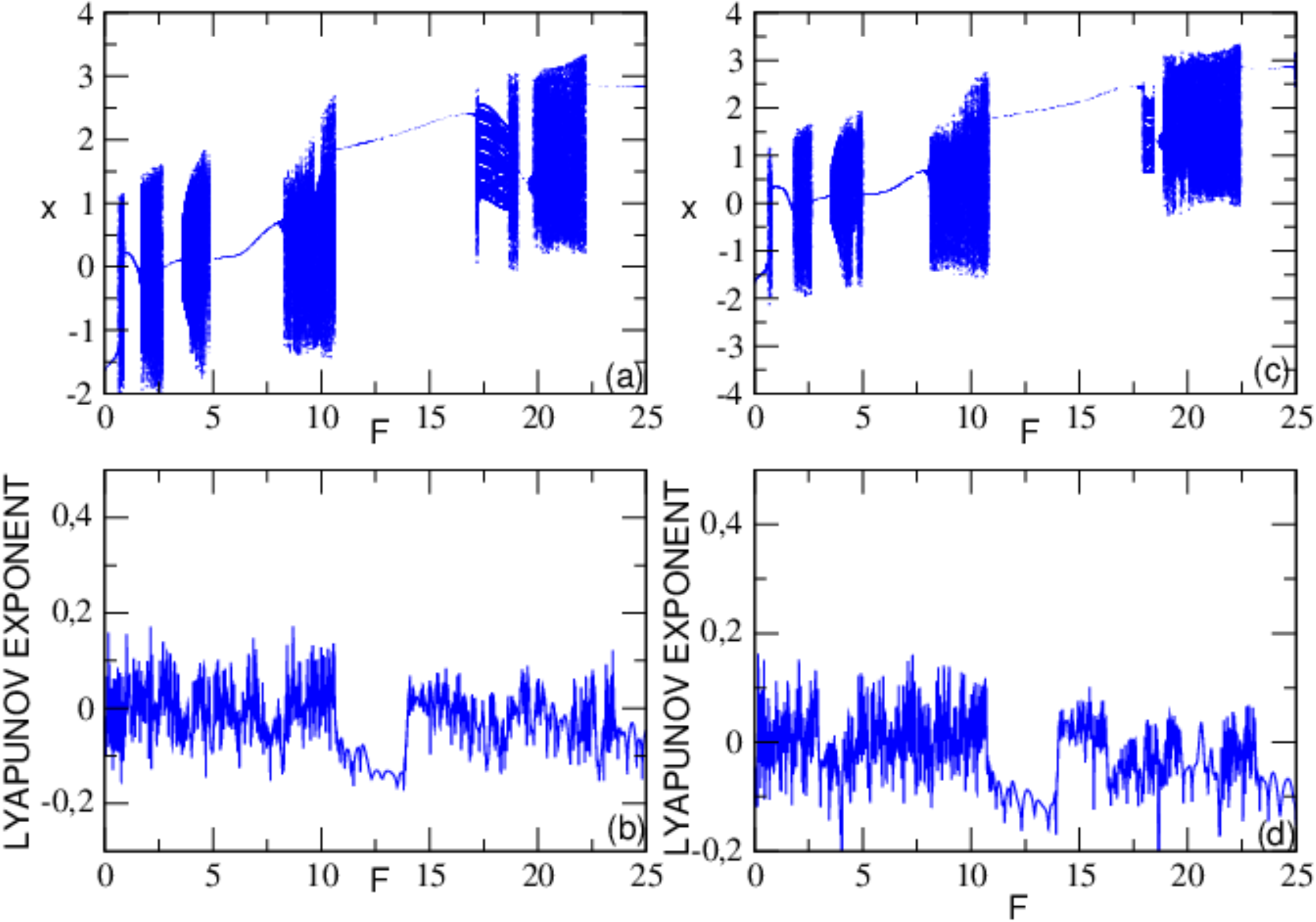}
\end{center}
\caption{Bifurcation diagram (upper frame) and Lyapunov exponent (lower frame) versus the amplitude 
$F$ with the parameters $\beta = 3.05; \gamma = 1.5; \Omega = 1;\epsilon = 0.02$; $\alpha=0 $ (left)
and  $\alpha=1$ (right).}
\label{fig:9}
\end{figure}

\begin{figure}[htbp]
\begin{center}
 \includegraphics[width=12cm,  height=10cm]{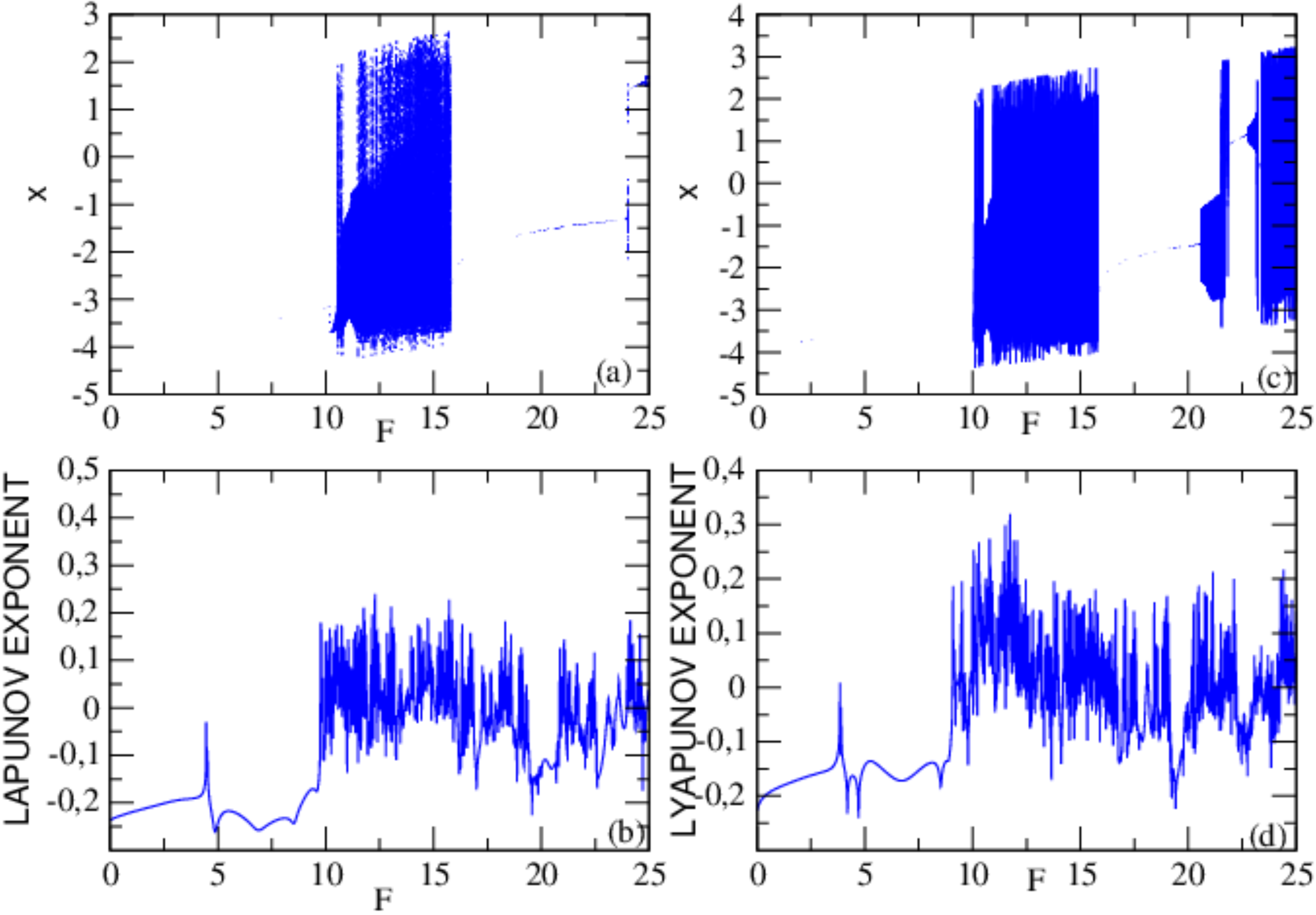}
\end{center}
\caption{Bifurcation diagram (upper frame) and Lyapunov exponent (lower frame) versus the amplitude 
$F$ with the parameters of Figure \ref{fig:9}for $\beta = 6$,  $\alpha=0 $ (left)
and  $\alpha=1$ (right).}
\label{fig:10}
\end{figure}

\begin{figure}[htbp]
\begin{center}
 \includegraphics[width=12cm,  height=10cm]{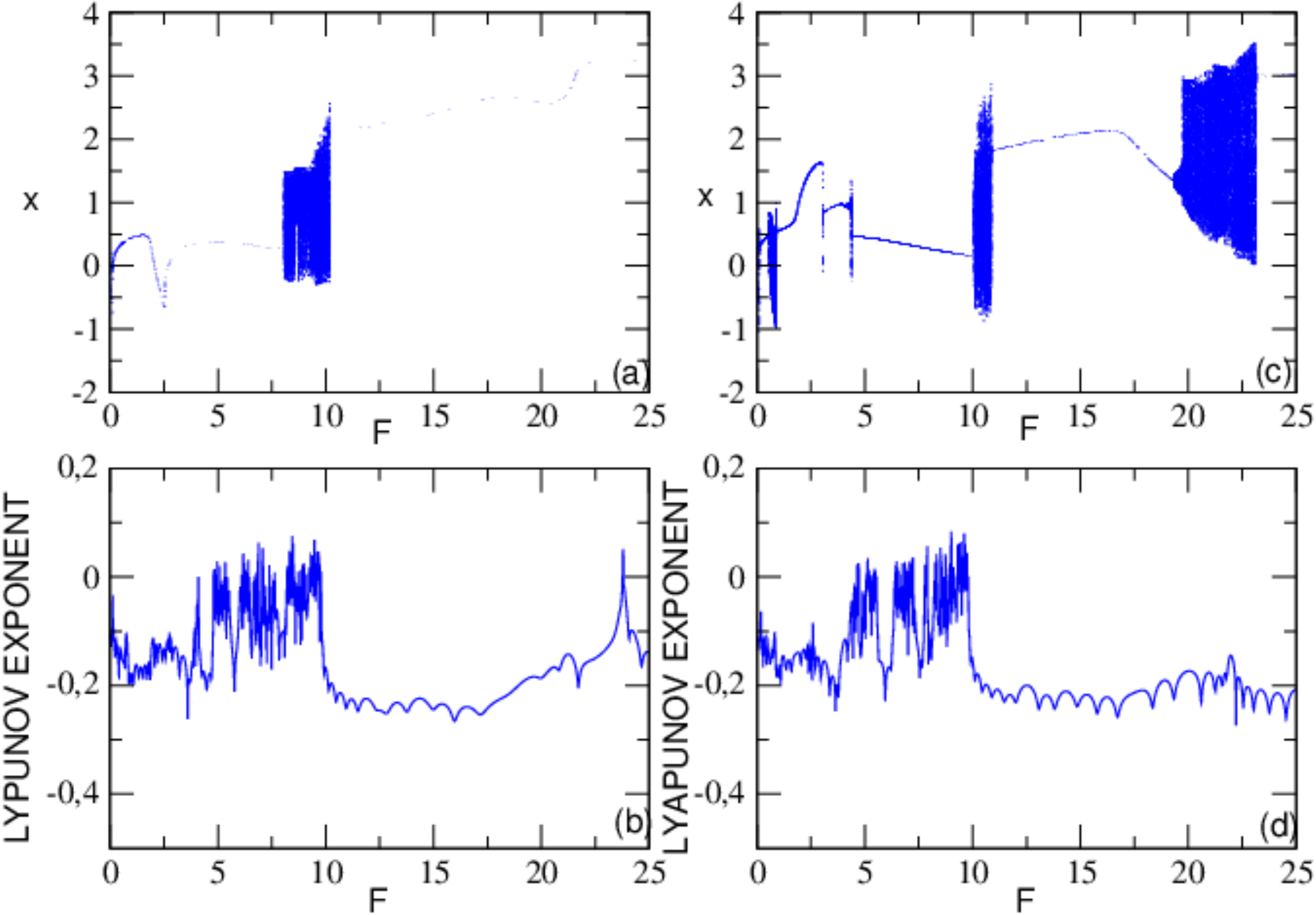}
\end{center}
\caption{Bifurcation diagram (upper frame) and Lyapunov exponent (lower frame) versus the amplitude 
$F$ with the parameters of Figure \ref{fig:9} for $\beta = 1.80; \epsilon = 0.06$,  $\alpha=0 $ (left)
and  $\alpha=1$ (right).}
\label{fig:11}
\end{figure}

\begin{figure}[htbp]
\begin{center}
 \includegraphics[width=12cm,  height=10cm]{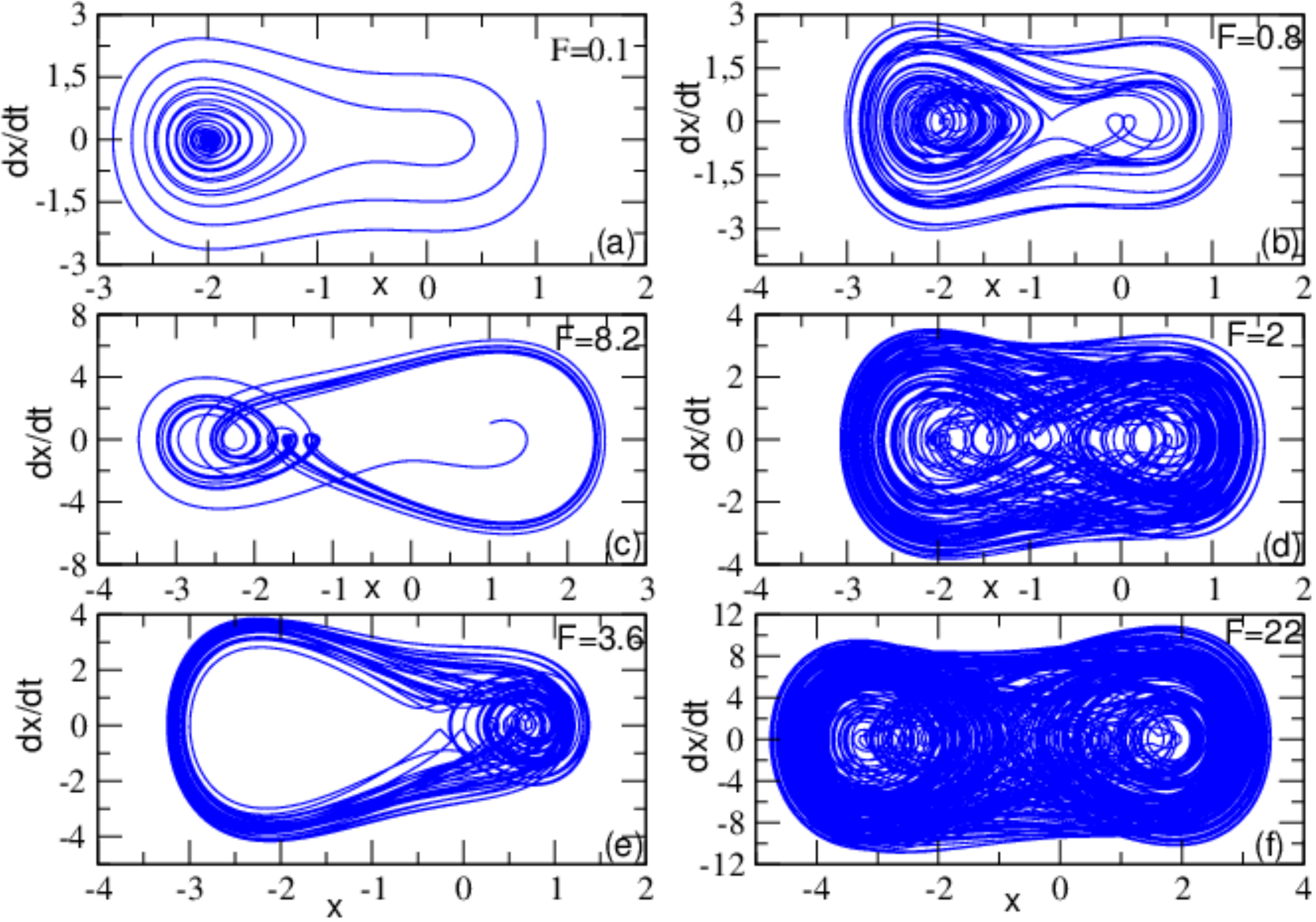}
\end{center}
\caption{Various phase portraits for several different values of $F$ with the
parameters of Figure 9 and $\alpha=0$.}
\label{fig:12}
\end{figure}

\begin{figure}[htbp]
\begin{center}
 \includegraphics[width=12cm,  height=10cm]{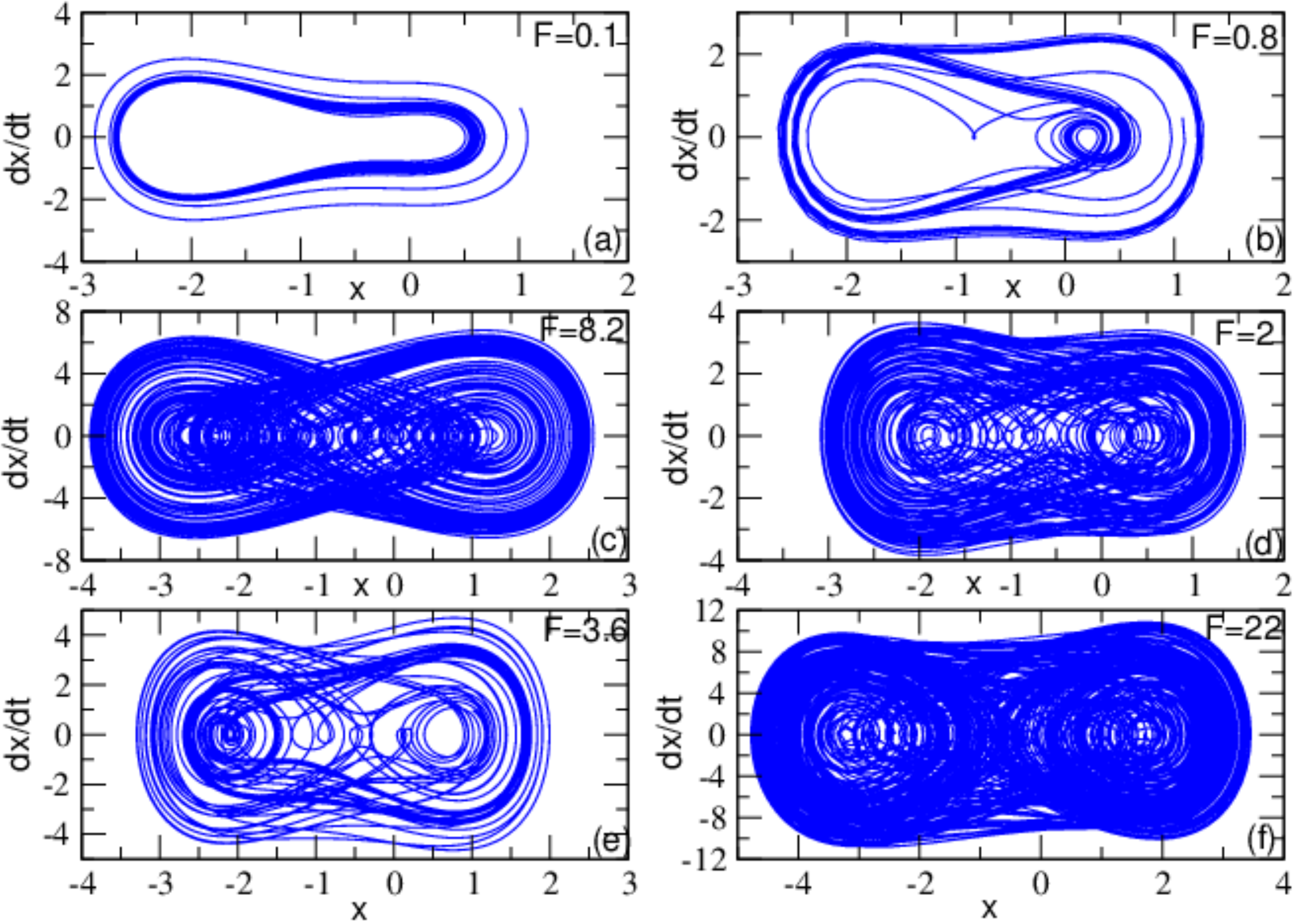}
\end{center}
\caption{Various phase portraits for several different values of $F$ with the
parameters of Figure \ref{fig:9} and $\alpha=1$.}
\label{fig:13}
\end{figure}

\begin{figure}[htbp]
\begin{center}
 \includegraphics[width=12cm,  height=10cm]{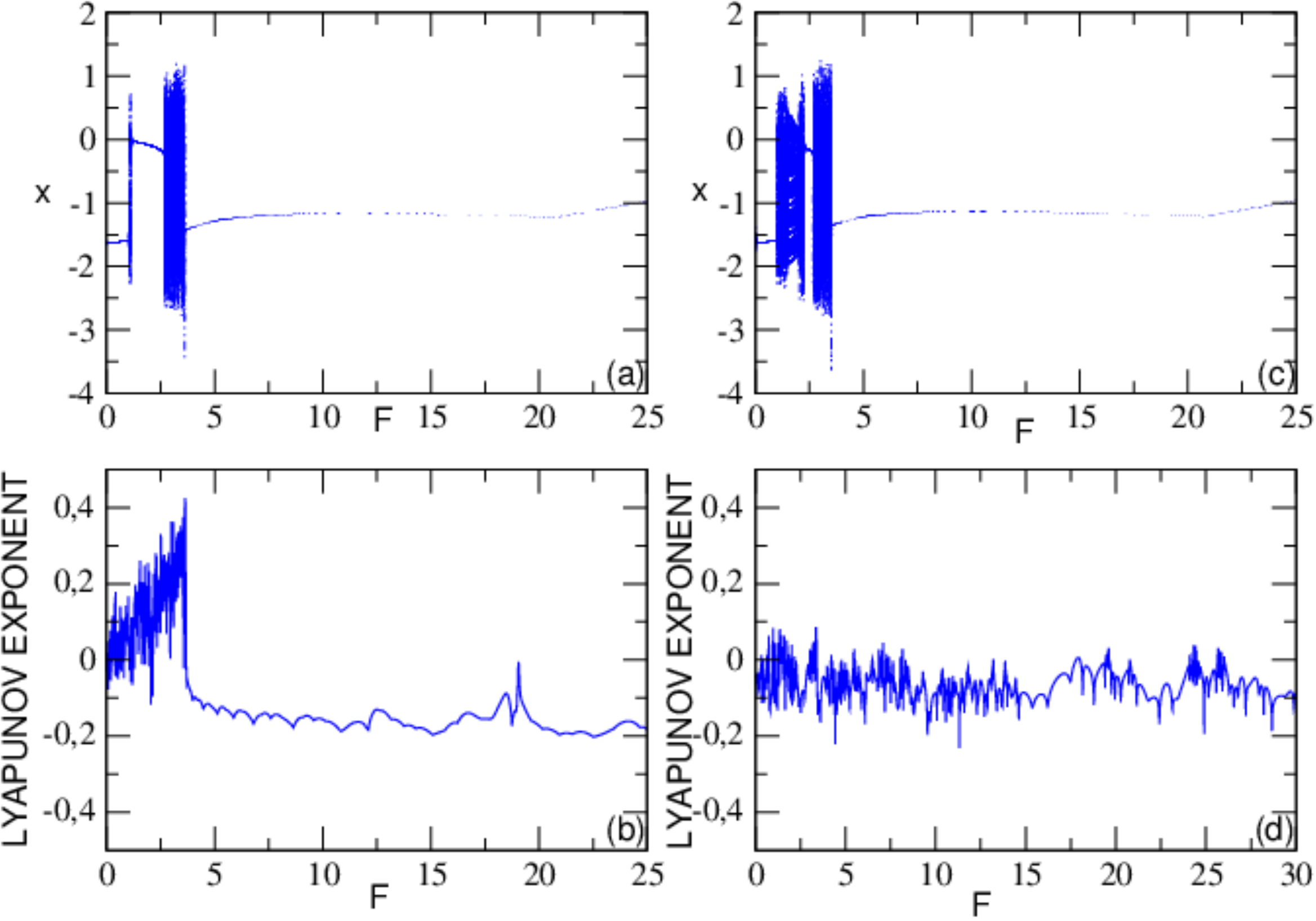}
\end{center}
\caption{Bifurcation diagram (upper frame) and Lyapunov exponent (lower frame) versus the amplitude 
$F$ with the parameters of Figure \ref{fig:9} for $\Omega=3 $,  $\alpha=0 $ (left)
and  $\alpha=1$ (right).}
\label{fig:14}
\end{figure}

\begin{figure}[htbp]
\begin{center}
 \includegraphics[width=12cm,  height=10cm]{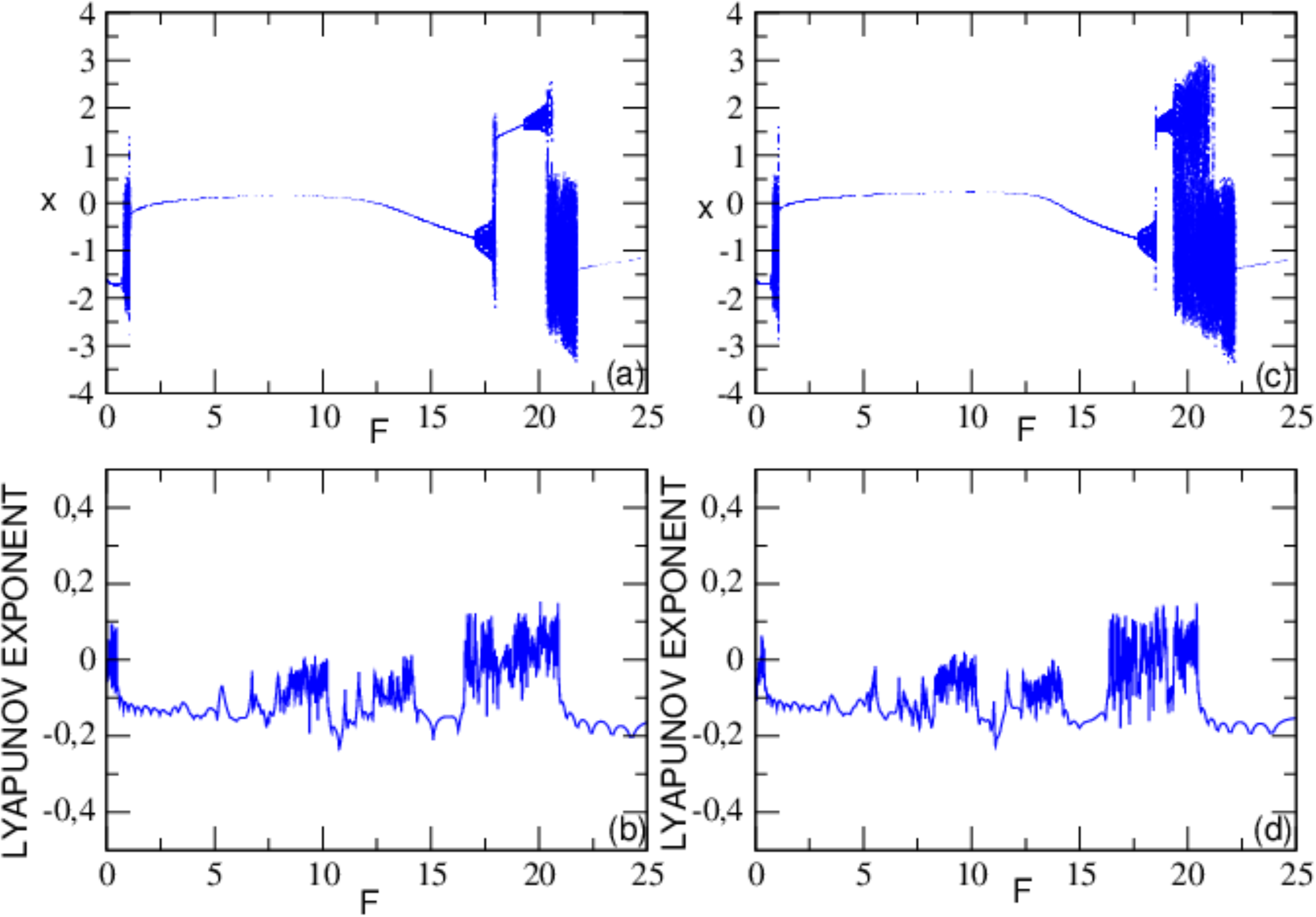}
\end{center}
\caption{Bifurcation diagram (upper frame) and Lyapunov exponent (lower frame) versus the amplitude 
$F$ with the parameters of Figure \ref{fig:9} for $\Omega=2$,  $\alpha=0 $ (left)
and  $\alpha=1$ (right)..}
\label{fig:15}
\end{figure}

\begin{figure}[htbp]
\begin{center}
 \includegraphics[width=12cm,  height=10cm]{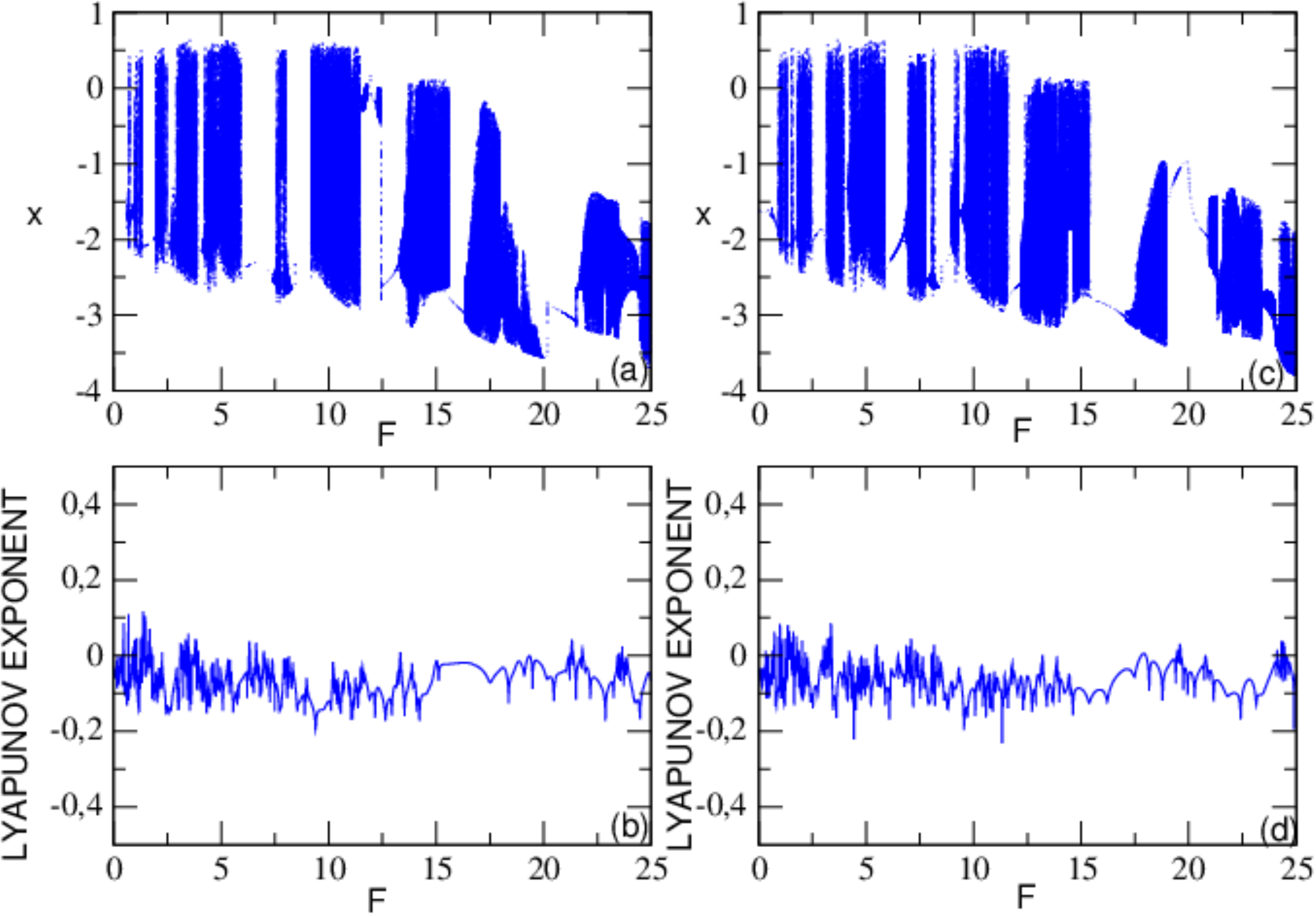}
\end{center}
\caption{Bifurcation diagram (upper frame) and Lyapunov exponent (lower frame) versus the amplitude 
$F$ with the parameters of Figure \ref{fig:9} for $\Omega=\frac{1}{3}$ ,  $\alpha=0 $ (left)
and  $\alpha=1$ (right).}
\label{fig:16}
\end{figure}

\begin{figure}[htbp]
\begin{center}
 \includegraphics[width=12cm,  height=10cm]{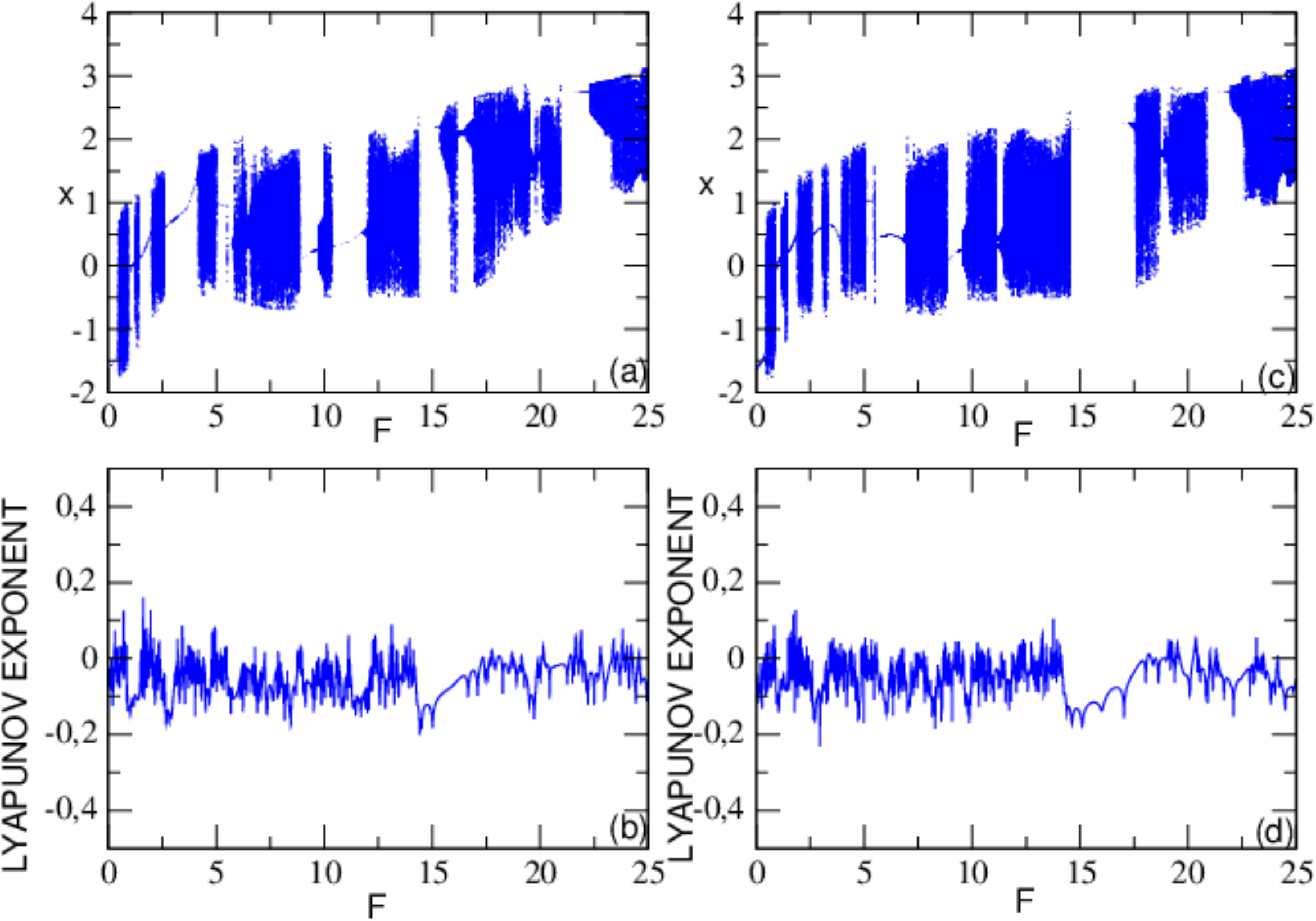}
\end{center}
\caption{Bifurcation diagram (upper frame) and Lyapunov exponent (lower frame) versus the amplitude 
$F$ with the parameters of Figure \ref{fig:9} for $\Omega=\frac{1}{2}$,  $\alpha=0 $ (left)
and  $\alpha=1$ (right).}
\label{fig:17}
\end{figure}

\begin{figure}[htbp]
\begin{center}
 \includegraphics[width=12cm,  height=8cm]{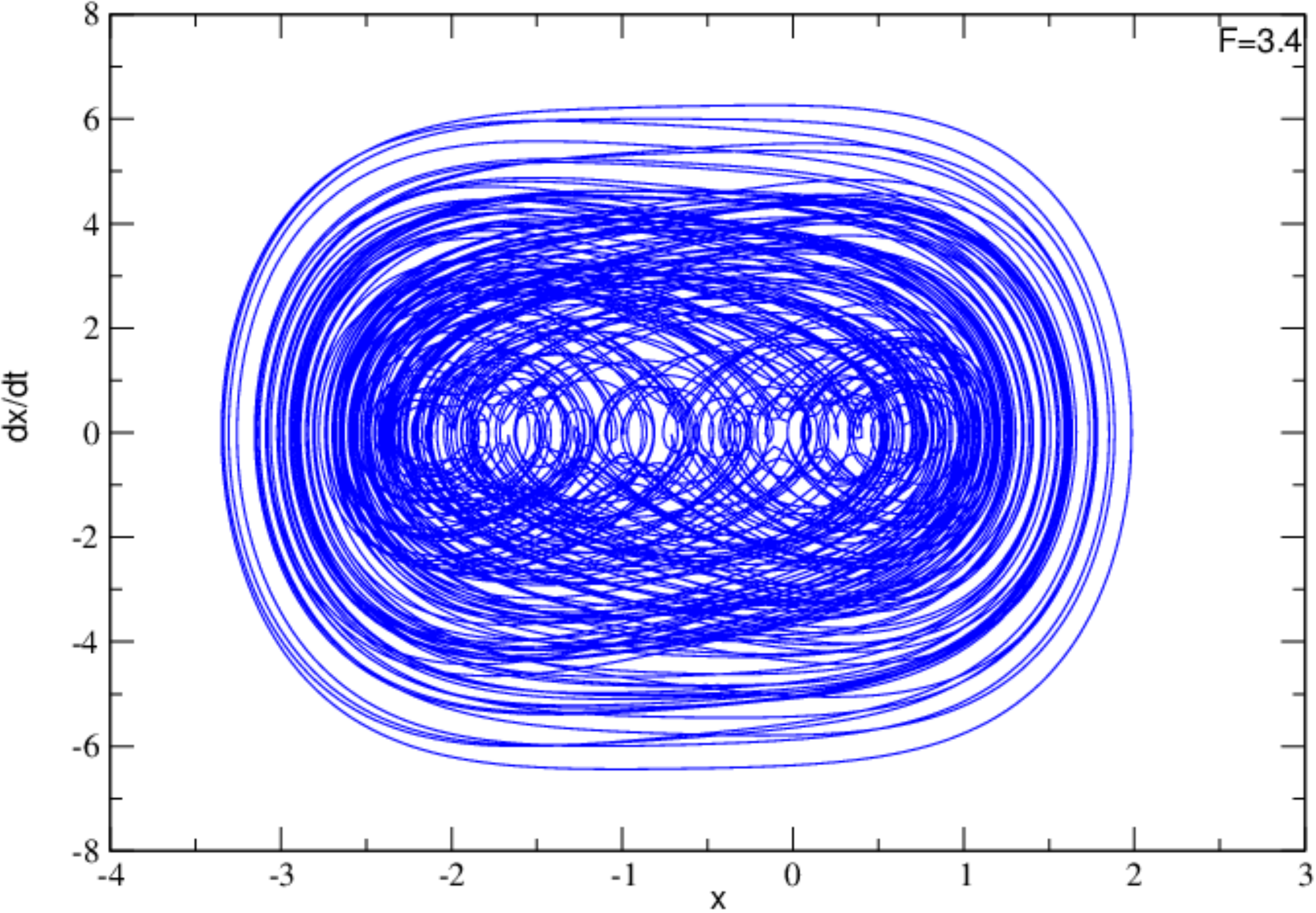}
\end{center}
\caption{Phase portrait for the first superharmonic resonance with the
parameters of Figure \ref{fig:14} }
\label{fig18}
\end{figure}

\begin{figure}[htbp]
\begin{center}
 \includegraphics[width=12cm,  height=8cm]{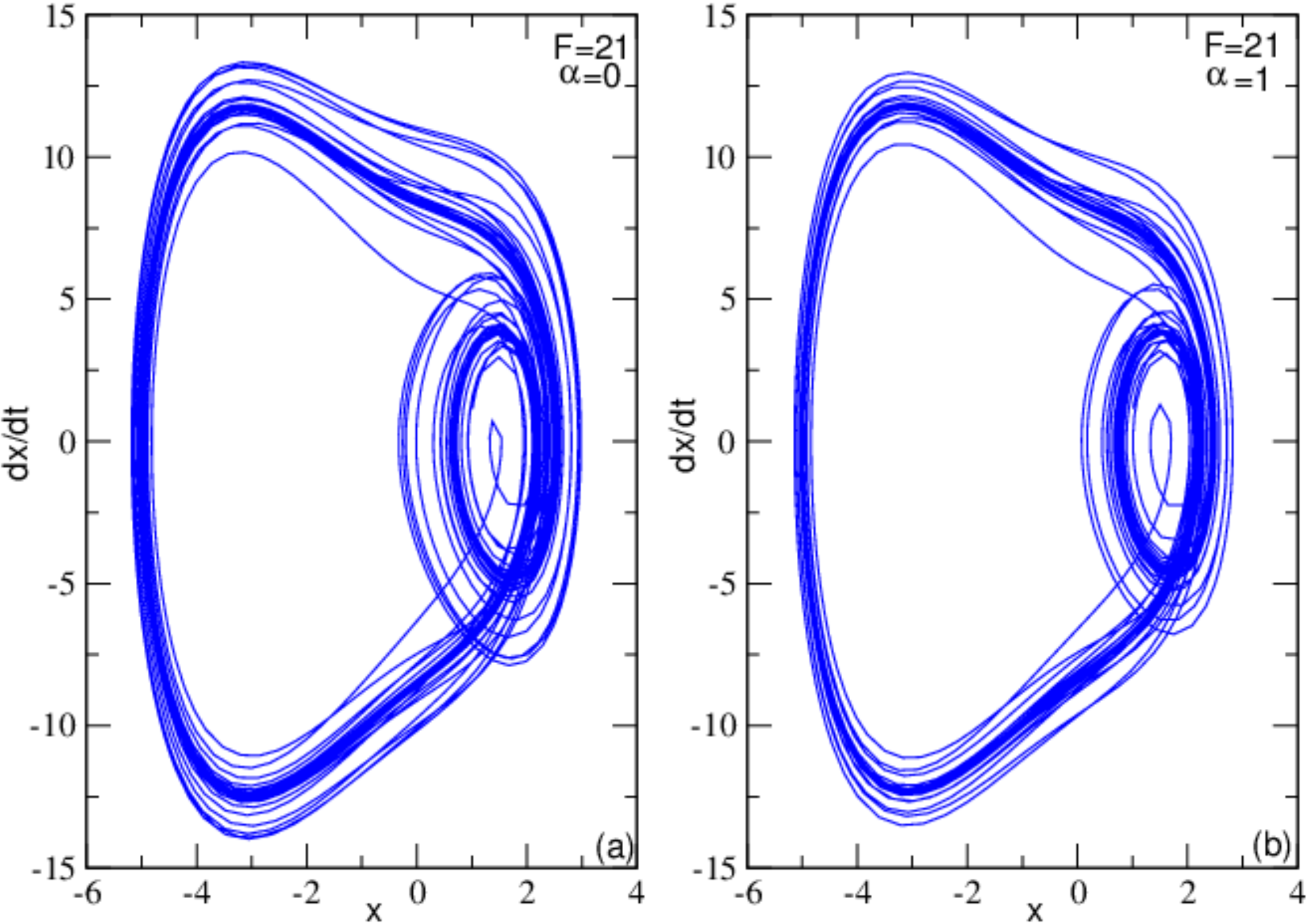}
\end{center}
\caption{Phase portrait for the second superharmonic resonance with the
parameters of Figure \ref{fig:15} .}
\label{fig:19}
\end{figure}

\begin{figure}[htbp]
\begin{center}
 \includegraphics[width=12cm,  height=8cm]{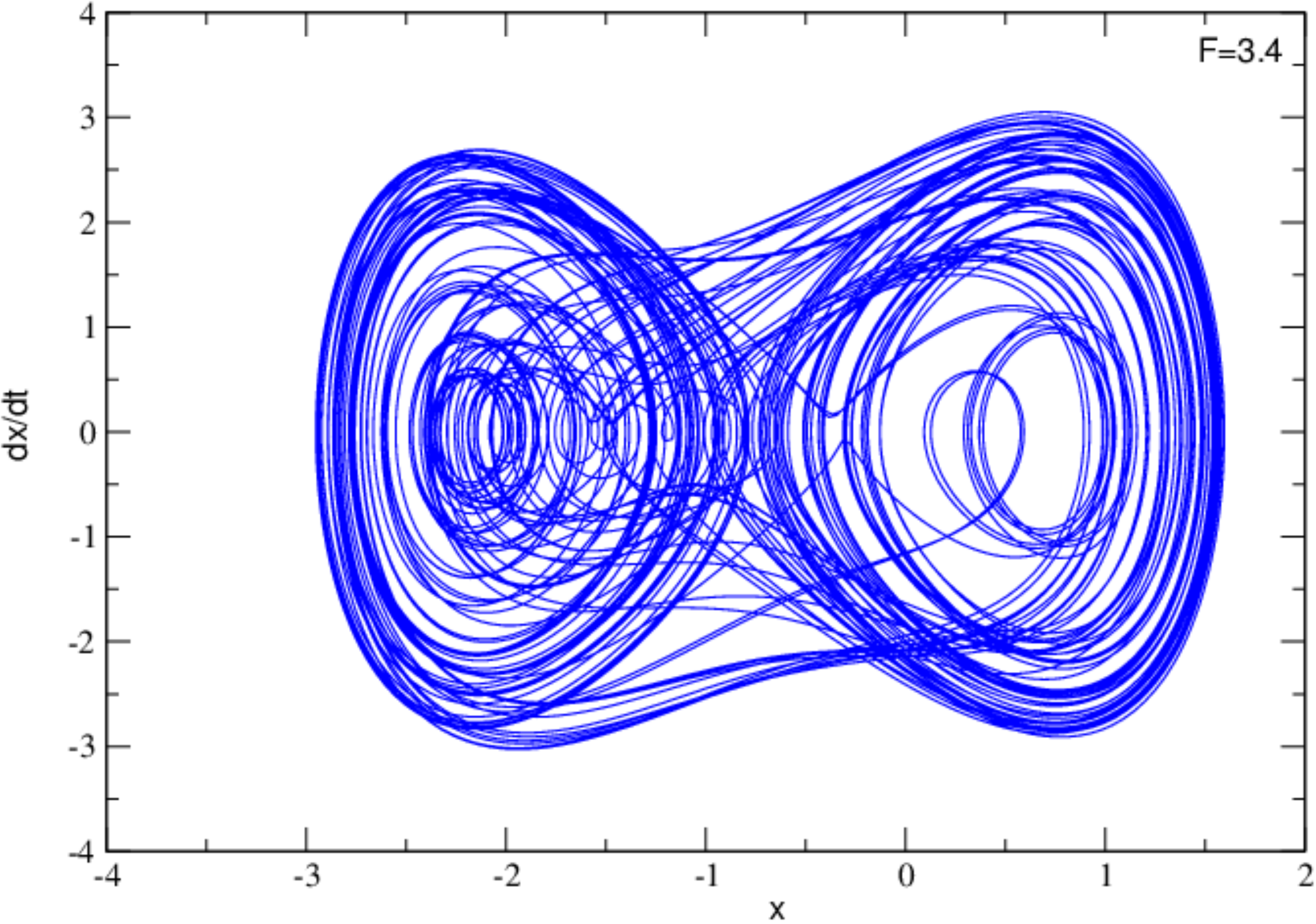}
\end{center}
\caption{Phase portrait for the first subharmonic resonance with the
parameters of Figure\ref{fig:16}.}
\label{fig:20}
\end{figure}

\begin{figure}[htbp]
\begin{center}
 \includegraphics[width=12cm,  height=8cm]{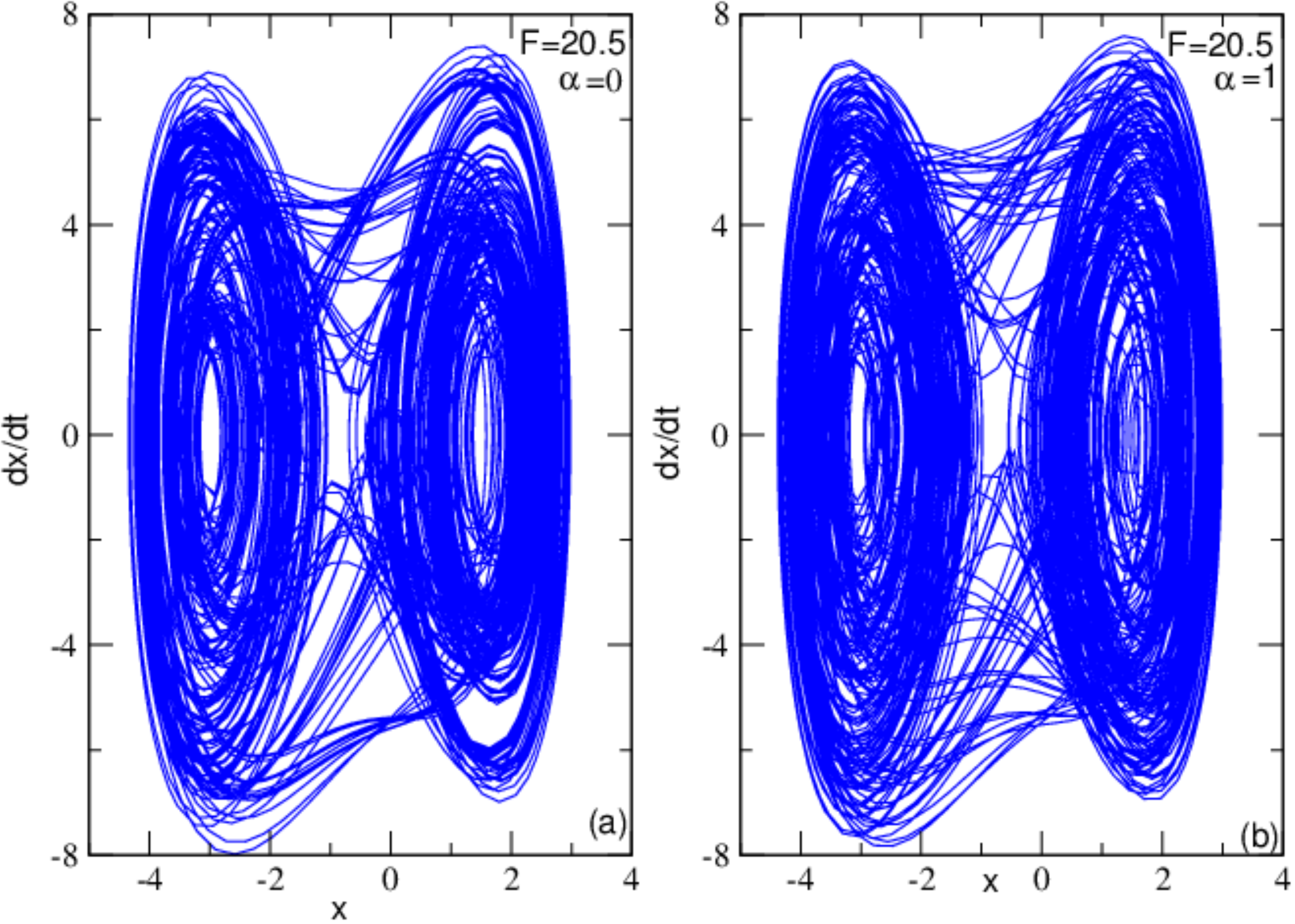}
\end{center}
\caption{Phase portrait for the second subharmonic resonance with the
parameters of Figure \ref{fig:17}.}
\label{fig:21}
\end{figure}

The aim of this section is to find some bifurcation structures in the nonlinear dynamics of plasma oscillations 
described by equation (\ref{eq20}) for resonant states since they are of interest in this system. For this purpose,  
we numerically solve this equation using the fourth-order Runge Kutta algorithm \cite{26} and plot the resulting bifurcation
 diagrams and the variation of the corresponding largest Lyapunov exponent as the amplitude $F$,  the parameters of nonlinearity 
$\epsilon,  \alpha,  \beta$ and $\gamma$ are varied. The stroboscopic time period used to map various transitions
which apper in the model is $ T=\frac{2\pi}{\Omega}$. \\
The largest Lyapunov exponent which is used here as the instrument to measure the rate of chaos in the system is defined as

\begin{equation}
 Lya=\lim_{t \rightarrow \infty}\frac{ln \sqrt{dx^2+d\dot x^2}}{t}\label{eq70}
\end{equation}
where $dx$ and $d\dot x$ are respectively the variations of $x$ and $\dot x$.
Initial condition that we are used in the simulations of this section is $(x_0, \dot x_0)=(1, 1)$.
For the set of parameters $\beta = 3.05,  \gamma= 1.5,  \epsilon = 0.02,  \Omega = 1$,  with 
$\alpha=0 $ (left) and $\alpha=1$ (right) the bifurcation diagram and its corresponding 
Lyapunov exponent are plotted in Fig. \ref{fig:9}. The same simulation are found in Fig.\ref{fig:10} and Fig. \ref{fig:11} with
respectively $\beta = 6$ and $\beta = 1.80; \epsilon = 0.06$.  From the bifurcation diagrams 
,  various types of motions are displayed. It is found that the model can switch from periodic to 
quasi-periodic oscillations or chaotic motions see Fig.\ref{fig:9}. In order to illustrate such situations,  we
have represented the various phase portraits using the parameters of the bifurcation diagram for 
which periodic,  quasi-periodic oscillations and chaotic motions are observed in Fig. \ref{fig:12} with effect
of parameter $\alpha$ in Fig.\ref{fig:13}. These observations prove that the model is highly to the initial
 conditions.  It should be emphasized from Fig.\ref{fig:9} that there are some domains where the Lyapunov exponent does not match very well the regime of oscil-
lations expected from the bifurcation diagram. Far from being an error which has occurred from
the numerical simulation process,  such a behavior corresponds to what is called the intermittency phenomenon. Therefore,  within these intermittent domains,  
the dynamics of the model can not be predicted. For instance,  some forecasted period-1 and quasiperiodic motions from the bifurcation
diagram are not confirmed by the Lyapunov exponent. From Figs. \ref{fig:12} and \ref{fig:13},  it is clearly that when the
quadratic hybrid parameter  the chaotic as well the ones of intermittency remain in the system. 
On the other hand,  the  parameters of nonlinearity can influence the chaotic motion in model.
Consequently a set of physical parameters of the model,  $\alpha,  \beta,  \epsilon$ and $\gamma$
can be used to increase or dismiss the rate of chaotic motion in the model. 
The same simulations are obtained in the subharmonic and superharmonic resonance have also been 
represented respectively in Figs. \ref{fig:14} - \ref{fig:21}.  From these figures,  we conclude that chaos is more abundant
in subharmonic resonant states than in the superhamonic and primary resonances.

\newpage

\section{Conclusion}
In this paper,  we have studied the nonlinear dynamics of plasma oscillations
modeled by forced modified Van der Pol-Duffing oscillator. The model has been described and the corresponding equation of motion obtained.
In the harmonic case,  the balance method has enabled us to derive the amplitude of harmonic oscillations and the effects of the differents
parameters on the behaviors of model have been analyzed. For the resonant states case,  the response amplitude,  stability have been derived
using multiple time-scales method and pertubation method it appears the two-order and three-order superharmonic and subharmonic resonances.
The effects of diffeerents parameters on these resonances are been found and we noticed that the hybrid and pure quadratic parameters have
several action on two-order resonances and three-order resonance are affected by the cubic nonlinear parameters. The influences of the 
dissipative parameter on the resonant,  hysteresis and jump phenomena have been highlighted. Our analytical results have
been confirmed by numerical simulation. Various bifurcation structures showing different types of transitions from quasi-periodic motions to
periodic and chaotic motions have been drawn and the influences of different parameters on these motions have been study. It is noticed 
that chaotic motions have been controlled by the  parameters $\epsilon,  \beta$ and $\gamma $ but also the hybrid quadratic parameter $\alpha$.
The results show a way to predict admissible values of the signal amplitude for a corresponding set of parameters. This could be helpful for 
experimentalists who are interested in trying to stabilize such a system with external forcing. Through these studies we notice that the hybrid 
quadratic therm have not neglected in study of nonlinear dynamics of plasma oscillations. For practical interests,  it is useful to develop tools
 and to find ways to control or suppress such undesirable regions. This will be also  useful to control high amplitude of oscillations obtained 
and which are generally source of instability in plasma physics.

\section*{Acknowlegments}
The authors thank IMSP-UAC and Benin gorvernment for financial support.


\begin{thebibliography}{100}
\bibitem{1}C. Hayashi,  
\emph{ Nonlinear Oscillations in Physical Systems}, 
(McGraw Hill,  New York,  $1964$),  
Sec.$ 1.5$ .
\bibitem{2} A. H. Nayfeh,  
\emph{ Introduction to perturbation techniques},  
(John Wiley and Sons,  New York,  $1981$), 
Sec.$4.5$.
\bibitem{3} J. Guckenheimer and P. J. Holmes,  
\emph{ Nonlinear oscillations,  
dynamical systems and bifurcations
of vectors fields},  (Springer-Verlag,  Berlin,  New York,  $1984$),  Sec. $1.3$.
\bibitem{4}S.
H. Strogatz,  
\emph{Nonlinear dynamics and chaos with applications to physics,  chemistry and engineering},  (Westview Press,  Cambridge,  $1994$),  Sec. $1.2$.
\bibitem{5} J.
D. Murray,  
\emph{Mathematical biology},  (Third edition,  Springer,  New York,  $2001$),  pp. $218-271$.
\bibitem{6} G. V. Paeva,  
\emph{Steath phenomena in dusty plasma},  (PhD thesis,  Technische Universiteit 
Eind-hoven,  $2005$).
\bibitem{7} S. Park,  C. R. Seon and W. Choe,  
\emph{Physics of Plasma} $11$,  $5095$ ($2004$).
\bibitem{8} R. A. Mahaffey,  
\emph{Physics of Fluids} $19$,  $1837$ ($1976$).
\bibitem{9} K. Ostrikov,  
\emph{Rev. Mod. Phys} $77$,  $489$ ($2005$).
\bibitem{10} K. Ostrikov and S. Xu,  
\emph{Plasma-aided Nanofabrication: from Plasma Sources to 
Nanoassem-bly} (John Wiley Sons,  Weinheim,  $2007$),  pp. $149-280$.
\bibitem{11} H. G. Enjieu Kadji,  J. B. Chabi Orou and P. Woafo,  
\emph{Physica Scripta} (In Press) ($2007$).

\bibitem{12}H. G. Enjieu Kadji,  B. R. Nana Nbendjo,  J. B. Chabi Orou,  and P. K. Talla, 
\emph{ Nonlinear dynamics of plasma oscillations modeled by an anharmonic oscillator} ($2007$).

\bibitem{13} B. E. Keen and W. H. Fletcher,  
\emph{Journal of Physics A: Gen. Phys} $5$,  $152$ $(1972)$.
\bibitem{14} B. E. Keen,  and W. H. Fletcher,  
\emph{Plasma Physics} $13$,  $419$ ($1970$).
\bibitem{15}L. H. Li and M. Matsuoka,  
\emph{Radiophysics and Quantum Electronics} $99$,  $N0.1$,  $75$ ($1996$).
\bibitem{16} J.Loverich and U. Shumlak,  
\emph{Physics of Plasma} $13$,  $082310$ ($2006$).
\bibitem{17}R. Bhattacharyya and M. S. Janaki, 
\emph {Physics of Plasma} $13$,  $044508$ ($2006$).

\bibitem{18} U. Shumlak and J. Loverich,  
\emph{Journal of Computational Physics} $187$,  $620$ ($2003$).
\bibitem{19}T. Kanki,  M. Nagata and T. Uyama,  $IEEE$ 
\emph{Transactions on Magnetics} Vol$42$,  $N0.4$,  $1403$
($2006$).
\bibitem{20}M. Arshad Mirza,  T. Rafiq and G. Murtaza,  
\emph{Physics of Plasma} $13$,  $1107$ ($1999$).
\bibitem{21}R. Dendy,  
\emph{Plasma Physics:an introductory course} (Cambridge University Press $1993$),  pp. $56$, 
$370$.
\bibitem{22}H. C. S. Hsuan,  
\emph{Phys. Rev} $172$,  $137$ ($1968$).

\bibitem{23} V. Land and W. J. Goedheer,  $IEEE$ 
\emph{Trans. Plasma Scie},  $35$,  $280$ ($2007$).

\bibitem{24} Rajasekar S,  Parthasarathy S,  Lakshmanan M. 
\emph{Prediction of horseshoe chaos in BVP and DVP oscillators}. Chaos,  Solitons and
Fractals $1992;2:271$.

\bibitem{25} J. Proud and al.,  
\emph{Plasma Processing of Materials: Scientific Opportunities and Technologies Challenges}. National Academy Press,  
Washington D.C. (1991).

\bibitem{26} N. Piskunov,  
\emph{Calcul Différentiel et intégral},  Tome II,  $9^e$ edition,  MIR,  Moscou (1980).


\end{thebibliography}
\end{document}